\documentclass[12pt]{iopart}
\usepackage{epsfig}
\usepackage{graphicx}
\usepackage{lscape}
\usepackage{longtable}

\begin{document}

\title[Young, nearby stellar associations]{The Tucana/Horologium, Columba, AB Doradus, and Argus Associations: New Members and Dusty Debris Disks }

\author{B. Zuckerman$^1$, Joseph H. Rhee$^1$, Inseok Song$^2$, M.S. Bessell$^3$,}

\address{$^1$Department of Physics and Astronomy, University of California, Los Angeles, CA 90095, USA}
\address{$^2$Department of Physics and Astronomy, University of Georgia, Athens, GA 30602-2451, USA} 
\address{$^3$Research School of Astronomy and Astrophysics, Australian National University, MSO, Cotter Road, Weston, ACT 2611, Australia}
\eads{\mailto{ben@astro.ucla.edu,rhee@astro.ucla.edu,song@physast.uga.edu, \\ bessell@mso.anu.edu.au}}
\begin{abstract}
We propose 35 star systems within $\sim$70 pc of Earth as newly identified members of nearby young stellar kinematic groups; these identifications include the first A- and late-B type members of the AB Doradus moving group and field Argus Association.  All but one of the 35 systems contain a bright solar- or earlier-type star that should make an excellent target for the next generation of adaptive optics (AO) imaging systems on large telescopes.  AO imaging has revealed four massive planets in orbit around the $\lambda$ Boo star HR 8799.  Initially the planets were of uncertain mass due in large part to the uncertain age of the star.  We find that HR 8799 is a likely member of the $\sim$30 Myr old Columba Association implying planet masses $\sim$6 times that of Jupiter.  We consider Spitzer Space Telescope MIPS photometry of stars in the $\sim$30 Myr old Tucana/Horologium and Columba Associations, the $\sim$40 Myr old field Argus Association, and the $\sim$70 Myr old AB Doradus moving group.  The percentage of stars in these young stellar groups that display excess emission above the stellar photosphere at 24 and 70 $\mu$m wavelengths  -- indicative of the presence of
a dusty debris disk -- is compared with corresponding percentages for
members of 11 open clusters and stellar associations
with ages between 8 and 750 Myr, thus elucidating the decay of debris disks with time.

\end{abstract}
\maketitle

\section{Introduction}

Young stars within 100 pc of Earth are excellent laboratories for the
study of stellar and planetary system evolution during ages from about ten
to hundreds of millions of years.  Excepting the Sun, it is 
difficult to deduce even a moderately accurate age of an individual star.  
Astronomers have circumvented this problem by studying stars
in rich open and globular clusters.  Excepting the Hyades, no such clusters
exist within 100 pc of Earth.  Fortunately, beginning in the late 1990s,
the existence of various substantial
co-moving, coeval, associations of young nearby stars has become evident (see reviews by
Zuckerman \& Song 2004 and Torres et al 2008).

Investigation of such stars will enhance knowledge of early stellar
evolution.  However, as a consequence of their proximity to Earth, in the
long term the most valuable contribution to astronomy of these youthful
stars will likely lie in the realm of the origin and early evolution of
planetary systems -- proximity buys one enhanced brightness and, for a
given linear scale, an enhanced angular scale.  Table 1 lists papers in which new members of the four nearest co-moving groups of young stars have been proposed previously; according to Torres et al. (2008) the mean distance from Earth is less than 50 pc for the most secure members of each of these four groups.    As new telescopes and cameras are being developed for imaging of young planets and the dusty
disks out of which they form, it is important to complete the inventory of
young stars near Earth.  

A principal motivation of the present paper is identification of new members of the AB Doradus and Tucana/Horologium Associations.  While considering the latter association we recognized new members of the Columba Association (Torres et al. 2008) including the now famous HR 8799 (= HIP 114189) that is orbited by at least four giant planets (Marois et al. 2010).  The age and some other properties of HR 8799 are considered in Section 5.3.

While considering the AB Dor Association we have identified its first known A- and late-B type members. In addition, we have identified field members of the Argus Association that are of earlier spectral type and generally nearer to Earth than members proposed previously by Torres et al (2008). 

The Infrared Astronomical Satellite (IRAS) discovered the phenomenon of
dusty debris disks in orbit around main sequence stars (Aumann et al
1984).  By now $\sim$150 main sequence stars within 120 pc of Earth appear to
have excess infrared emission detectable with IRAS.  Almost all of these
excesses are measured most convincingly at 60 microns wavelength (Moor et al 2006, Rhee et al
2007a and references therein), although occasionally instead at 25 microns
(Rhee et al 2007b and Melis et al 2010 and
references therein).  While the Infrared Space Observatory (ISO) added a few new debris disk stars (e.g., Habing et al. 2001, Silverstone 2000), Spitzer Space Telescope surveys provided the next major advance; open clusters, nearby stellar associations, and nearby field
stars have all been studied.  Spitzer surveys included a variety of goals, 
for example an understanding of the evolution of the quantity
and temperature of the dusty debris with time, the association of debris
disks with stars of different spectral classes, and association or lack
thereof of debris with known planetary and stellar secondaries.   IRAS was an all-sky survey whereas ISO and Spitzer pointed at only a small fraction of the sky.  They were followed by the AKARI all-sky survey that, typically, was no more sensitive than IRAS.  Most recently, the Wide Field Infrared Survey Explorer (WISE) has performed a mid-infrared all-sky survey far more sensitive than those of IRAS or AKARI.

A focus of the present paper is the evolution of debris disks as a
function of their age.  We present published and unpublished Spitzer data
for Tucana/Horologium, Columba, AB Doradus, and Argus Association members, identifying some
previously unrecognized dusty systems.  Then we compare the overall
debris disk status for these associations with previously published Spitzer results for
other stellar associations near Earth.  

\section{Sample Selection}

Memberships of the four most prominent young stellar associations closest to Earth -- $\beta$ Pic, TW Hya, AB Dor and Tuc/Hor -- are considered in papers listed in Table 1.  In the present paper we address the latter two, specifically we search for new members and for dusty debris disks.  Although stars of the Columba Association are generally more distant from Earth than those of the four Table 1 Associations (Torres et al 2008), because of various similarities between Tuc/Hor and Columba stars we searched for Columba stars within 65 pc of Earth.  The Argus Association, especially IC 2391, as proposed by Torres et al (2008; their Tables 11 \& 12) is on average much more distant from Earth than are the four Table 1 associations, but a few of their proposed Argus field members are within 70 pc of Earth.  This inspired us to search for additional nearby Argus field stars.  Proposed new members of AB Dor, Tuc/Hor \& Columba, and Argus, all within $\sim$70 pc of Earth, may be found in Tables 2-4.

Potential new members of these associations are identified by a match of ages and Galactic space motions UVW.  These characteristics can be measured or estimated via optical spectroscopy and astrometry and, also, optical and X-ray photometry.  Optical data for late-F through early-M type stars come principally from our spectroscopic survey described in Section 3.  For these spectral types we generally observed only stars that appear in the ROSAT All Sky Survey (RASS; Voges et al 1999, 2000).  For earlier spectral types an important source of data is the catalog of Gontcharov (2006) which includes accurate radial velocities for stars from late-B through early-F type.

For dusty debris disk studies with the Spitzer Space Telescope, MIPS photometry was obtained via our GO program $\#$3600 "Disk Census of Nearby Stellar Groups", but also via the Spitzer archives.  When poring through the Spitzer literature we noticed that refereed papers had been devoted to most young and/or nearby stellar clusters and associations except that AB Dor and Tuc/Hor were conspicuous by their absence; we are now filling this void.  Because some likely members of AB Dor and Tuc/Hor are identified in papers that post-date the cold Spitzer mission (including some papers listed in Table 1 and also the present paper) infrared photometry for these two stellar associations is incomplete.  Similar remarks pertain to Columba and Argus Association members. Perhaps WISE, Herschel, and the Stratospheric Observatory for Infrared Astronomy (SOFIA) will help to complete the disk census for stars in young, nearby associations.

\section{Observations and analysis}

For southern hemisphere stars spectra were obtained with the 2.3 m telescope at the Siding Spring Observatory (SSO) of the Australian National University.  Double-beam grating (DBS) and echelle spectrographs were employed at the two Nasmyth foci.  The red channel of the DBS covers the spectral range 6500 - 7450 \AA\ at a measured resolution of 1.16 \AA\ (0.55 \AA\ pixel$^{-1}$).  Eight orders of the echelle spectra cover portions of the wavelength range from 5800 to 7230 \AA.  We focus on orders that contain the H$\alpha$ and Li $\lambda$6708 lines.  In these orders the measured resolution was 0.45 \AA\ (0.17 \AA\ pixel$^{-1}$).  Radial velocities were determined by cross-correlating target and radial velocity standard spectra over 5 or 6 orders of the echelle that were chosen to produce strong correlations and have few atmospheric features.  
All spectra were reduced following a standard procedure (bad pixel and cosmic ray removal, flat fielding, source extraction, telluric correction, etc.) using IRAF.  Equivalent widths of H$\alpha$ and Li $\lambda$6708 were measured with the IRAF task splot.  

For northern hemisphere stars we used the Hamilton echelle spectrograph at the coude focus of the 3 m telescope at Lick Observatory.  Data reduction procedures were similar to those employed at Siding Spring.

As noted in Section 2, MIPS photometry was obtained via GO program $\#$3600 and also via the Spitzer archives.  For the MIPS 24 $\mu$m band, we performed aperture photometry on post-BCD
images provided by the Spitzer Archive using an aperture radius of 13" and sky
annuli of 20" and 32".  An aperture correction of 1.17 was applied based on
Table 4.13 of the MIPS Instrument Handbook version 2.0.  In cases where the target is
a binary object (e.g., HIP 116748), we used a larger aperture size to collect the combined Spitzer
flux densities and compared them with combined optical and near-IR flux densities.   For the MIPS 70 $\mu$m band, we first created a mosaic
image by combining BCD images with the Spitzer MOsaicker and Point source
EXtractor (MOPEX) tool and performed aperture photometry on the mosaic image.
For a few stars Spitzer IRS data exist and we used these where appropriate to confirm or deny the existence of excess IR emission.   Spitzer results are summarized in Tables 5-11.  Figures 1-7 display the spectral energy distributions (SED) of all AB Dor, Tuc/Hor, nearby Columba, and nearby Argus field stars that we deem to definitely or probably have excess infrared emission.   

A fully automated SED fitting technique employing a theoretical
atmospheric model (Hauschildt et al. 1999) was used to predict
stellar photospheric fluxes.  A detailed description of our photospheric fitting procedure is given in Section 2 of Rhee et al. (2007a), so we do not repeat it here.  To check whether our atmospheric model well reproduces actual stellar photospheres, in Figure 8 we compare the model predictions with MIPS measurements at 24 $\mu$m of a sample of nearby F, G, and K-type stars from a paper by Trilling et al. (2008).  Stars plotted in Figure 8 do not include any stars deemed by Trilling et al to have either 24 or 70 $\mu$m excess emission (their Tables 5 and 6).  The plotted stars are all 5th magnitude or fainter at K-band as measured in the 2MASS survey.  We avoid brighter stars that may be too bright to be measured accurately by 2MASS in J-band.
As may be seen in Figure 8, for K, G, and F-type stars the MIPS measured 24 $\mu$m fluxes are on average about 3$\%$ larger than the model predicted fluxes.  While some of this offset may be due to small and previously unrecognized excess IR emission from some stars in the Trilling et al sample, conservatively, we ignore this possibility and treat the 3$\%$ as an effective error in the photospheric model predictions.  Therefore, in Tables 8-10 the listed 24 and 70 $\mu$m photospheric fluxes are those predicted from the theoretical atmospheric model but multiplied by a factor of 1.03.  In addition to this small mean offset, the Figure 8 histogram indicates a $\pm$7\% uncertainty in the individual 24 $\mu$m flux densities due to some combination of model uncertainties, measurement error, and perhaps other factors. 

The MIPS Handbook lists 4$\%$ as the systematic 24 $\mu$m absolute flux calibration uncertainty (see also Engelbracht et al. 2007).  This 4$\%$ is negligible when added in quadrature with the measurement uncertainty estimated just below.        

We checked the 7$\%$ flux density uncertainty deduced from Figure 8 in the following fashion.  For stars in Tables 8-10 with measured 24 $\mu$m flux density less than the model photosphere (corrected upward by the factor of 1.03 mentioned just above) we calculate a percentage deficit given by the ratio of photosphere to measured 24 $\mu$m flux density and then minus unity.  For K8 to M3 stars in the AB Dor group (Tables 5 and 8) this flux deficit often falls between 14 and 33$\%$. But for earlier-type AB Dor stars the largest flux deficits are more modest;  only HIP 19183 (14\%) and HD 317617 (15.5$\%$) have flux deficits $>$11$\%$.  (HD 317617 is the faintest 24 $\mu$m star in Table 8.)  In Table 9 (Tuc/Hor and Columba stars) the only non-M-type stars with flux deficits $>$11$\%$ are HIP 2484 (14$\%$), HIP 2487 (15.5$\%$), and CD-34 2406 (23\%).  (The latter star is by far the faintest 24 $\mu$m star listed in Table 9.)   

The preceding paragraph indicates that while $\pm$11\% appears to be an appropriate 24 $\mu$m flux density uncertainty for most non-M-type stars in Tables 8-10, adaptation of 16\% for this uncertainty should encompass just about  all such stars.  This choice also encompasses the 7$\%$ flux uncertainty at 24 $\mu$m for stars in Fig. 8, as mentioned above.

Thus, when a MIPS measured 24 $\mu$m flux is a factor 1.16 larger than a photospheric flux listed in Tables 8-10, we deem a non-M-type star to have excess flux at 24 $\mu$m.  If the apparent 24 $\mu$m MIPS excess lies between 12 and 16\%, then in the fourth column of Tables 8-10 we enter Y? indicating possible excess emission.  In fact, only 3 stars in Table 9 have entries of Y?, and only the binary star HIP 16563 (considered in Section 5.5.1) has a Y? entry in Table 8.

At 70$\mu$m, the procedure is similar, except that we use 7$\%$ as the calibration uncertainty as recommended in the MIPS Handbook.
\vskip 0.1in

\section{Results}

Tables 2-4 present stars we propose as new members of the AB Doradus, Tucana/Horologium, Columba, and field Argus Associations within $\sim$70 pc of Earth.  Based on their survey of southern hemisphere stars, Torres et al (2008) introduced the Columba and Argus Associations.  In addition to location in the southern hemisphere, all Columba members listed by Torres et al. lie in the R.A. range between about 2h and 8h.  In our study we have placed no a priori restriction on R.A. or Decl. for stars in any moving group.  Proposed membership is based on UVW and standard age indicators (see below and Section 5).   Future observations and traceback analysis can confirm or deny membership of each star.

According to Torres et al (2008), the U and W components of Columba's space motions are more negative than those of Tuc/Hor (see Section 5.2 and the notes to Table 3).  Also, according to Table 2 of Torres et al, Columba stars are more distant, on average, than stars in Tuc/Hor.  The listed UVW of a few stars in our Table 3 make it difficult to place them cleanly into either Tuc/Hor or Columba.   Only a few Columba stars in Table 5 of Torres et al are as close to Earth as stars listed in our Table 3.  Nine stars in our Table 3 lie in the northern hemisphere, whereas all but one of the Tuc/Hor and Columba stars in Zuckerman \& Song (2004) and Torres et al (2008) lie in the southern hemisphere.   Since Tuc/Hor and Columba have similar ages (30 Myr, Torres et al 2008), for the purposes of understanding the evolution of dusty debris disks as a function of time, in Table 11 and Section 5.5.2 we consider members of these two associations together.
    
As in our previous papers on these and other young, nearby,
moving groups (e.g., Song et al 2003, Zuckerman et al 2004, Zuckerman et
al 2001a and 2001b), we rely on a combination of techniques -- including
Galactic space motions UVW, location on color-magnitude diagrams, lithium
abundance, and X-ray luminosity -- to establish age and membership.   

Tables 5 and 6 list Tuc/Hor and AB Dor stars observed by
Spitzer as part of our GO program $\#$3600 or as part of Spitzer programs by other
groups; Table 7 lists these various programs for the proposed Argus Association members within 70 pc of Earth. Tables 8-10 present MIPS photometry while
Table 11 compares the frequency of dusty debris disks in the four moving groups considered in the present paper with that of other stellar associations with previously published Spitzer results.

\section{Discussion} 

\subsection{AB Doradus} 

While published compendia of members of the nearby Tucana/Horologium and $\beta$
Pictoris kinematic groups include stars of A-type or late B-type or both,
no stars earlier than mid-F spectral class are listed previously for the
AB Doradus moving group (Zuckerman et al. 2004; Zuckerman \& Song 2004; Torres et al
2008;  da Silva et al 2009).  Stars proposed in Table 2 as AB Dor members
are either of A- or late B-type.  Given that AB Dor stars appear to be
about as numerous as Tucana and $\beta$ Pictoris stars, identification of
these early-type AB Dor members while belated, is not unexpected.  All 
Table 2 stars lie near the location of a typical Pleiades star on an A-star color-mag diagram (Figure  5 in Zuckerman 2001) consistent with the age of the AB Dor moving group.

HIP 22845 (HD 31295) is a well studied $\lambda$ Boo and Spitzer and IRAS infrared excess star.  Its UVW of -4.3, -23.6, -10.1 $\pm$(1.1, 0.8, 0.5) km s$^{-1}$ is, at best, in marginal agreement with published UVW for the AB Dor group (see Note to Table 2).  HIP 22845 lies near typical Pleiades stars on an A-star color-mag diagram.  We have not included it as a proposed AB Dor member.

HIP 117452 (Table 2) is a member of a triple system. The A0 primary is a close binary and the tertiary, HD 223340, is an early K-type star about 75" to the NW.  Both the (binary) primary and HD 223340 are X-ray sources.   In the K-star the lithium line equivalent width = 148 m\AA.

We regard HIP 93580 (Table 2) as only a possible member of the AB Dor group because the stellar UVW is somewhat discordant with that of the mean UVW for the group (as may be seen from the entry in Table 2 and the Note to the table).

In addition to the proposed Table 2 early-type additions to the AB Dor group,
our spectroscopic studies at SSO indicate that solar-type stars HD 293857,
UX Col, and HD 178085 are also members (as noted independently by Torres et
al 2008 and da Silva et al 2009).

\vskip 0.2in

\subsection{Tucana/Horologium and Columba}

The Tucana and Horologium Associations were proposed independently by
Zuckerman \& Webb (2000) and Torres et al (2000), respectively.  Zuckerman
et al (2001b) suggested that these two moving groups are really just two
adjacent regions that contain a coeval ($\sim$30 Myr old) stream of stars with
common space motion, but that the greatest concentration of stars, the
"nucleus" of the overall group, is located in Tucana.  Subsequently,
Torres et al (2008) introduced the notion that there are three 30 Myr old
associations, Tuc/Hor, "Columba", and "Carina".  As defined by Torres et al. (2008), these three differ principally in location in space -- both in the plane of the sky and in
distance from Earth -- and, to a lesser degree, probably in Galactic space
motions UVW (see also Section 4 and the note to Table 1).  

Some stars in Table 3 that are likely Columba members represent a major departure from the characteristics of the Columba stars listed by Torres et al. (2008); their listed stars all fall in the R.A. range between about 2 and 8 hours and, with but one exception at +4 deg, all have negative declinations.  (The latter regularity should not come as a surprise since the Torres et al study focused on the southern hemisphere.)  Also, only 5 of the 41 Columba stars they list are within 65 pc of Earth.  By contrast all 14 of the likely Columba members of our Table 3 (based on their large negative W component) are within 65 pc.  In addition, 7 of these 14 lie well outside of the plane of the sky boundaries (in R.A. and Decl.) that encompass all 41 of the proposed Torres et al. members.

While the Columba Association stars proposed by Torres et al. (2008) lie primarily between about 2h and 7h in R.A. and in the southern hemisphere, they are only weakly constrained in distance from Earth, including even a star 189 pc away.  We suggest that it would be preferable to constrain the membership to lie much closer to Earth, say out to $\sim$80 pc, and at all plane of the sky locations.  An 80 pc radius includes most of the members of the $\beta$ Pic, Tuc/Hor, AB Dor groups as proposed by Torres et al. (2008) and by Zuckerman \& Song (2004).   Based on Table 5 of Torres et al (2008) and our Table 3, the nucleus of such a Columba moving group would lie between 5h and 6h R.A. and would be about 50 pc from Earth (cf. all the stars in Table 3 from HIP 23179 to 32104, inclusive).

The F0 star HIP 17675 (HD 23384) does not appear in Table 3 although it is young and has a UVW consistent with that of Columba stars.  With a radial velocity of -1.6$\pm$0.7 km s$^{-1}$ (a weighted average of our echelle measurement and that given in Gontcharov 2006), the UVW of HIP 17675 is -11.3, -21.8, -5.7 $\pm$(0.9, 1.1, 0.4) km s$^{-1}$.  On an A-star color-magnitude diagram (Figure 5 in Zuckerman 2001), the F0-type primary lies low thus suggesting a young age.  However, our measurement of the equivalent width (EW) of the 6708 \AA\ line of lithium is only 50 m\AA\ and the  fractional X-ray luminosity, log(L$_x$/L$_{bol}$) based on the ROSAT All-Sky Survey (RASS), is about -6.0.  Together the lithium EW and the X-ray luminosity may be too small to be consistent with an age as young as 30 Myr.

Another interesting F0 star that did not find its way into Table 3 is HIP 82587 (HD 152598).   The UVW of -10.7, -22.9, -3.6 $\pm$(0.4, 0.5, 0.5) km s$^{-1}$ is similar to those of the Tuc/Hor and Columba Associations.   The star has MIPS measured excess emission at both 24 and 70 $\mu$m (Moor et al. 2009).  Moor et al. give an age of 210 $\pm$70 Myr based on the star's UVW, X-ray flux, and 6708 \AA\ lithium line strength.  Our interpretation of these characteristics is consistent with an age younger than 210 Myr, although perhaps not as young as 30 Myr.  Also, like HIP 17675 above, HIP 82587 lies quite low on the A-star color-mag diagram, suggestive of a young age.  Hence, membership in the Tuc/Hor or Columba Association remains a possibility.

In addition to stars listed in Table 3, based on our independent analysis, we concur with the classification by Torres et al. (2008) of HD 38206 (HIP 26966) as a member of the Columba association.   And we note that, consistent with the fact that our Table 3 stars are generally closer to Earth than the Columba stars proposed by Torres el al, HD 38206 at 69 pc from Earth is one of their nearer members. 

Comments on some of the stars in Table 3 follow next.  The quantity f$_x$ is the fractional X-ray luminosity, log(L$_x$/L$_{bol}$) based on the ROSAT All-Sky Survey (RASS).  The quoted lithium line equivalent widths (EW) are those of the 6708 \AA\ line.
\vskip 0.2in
\noindent HIP 12413: this is a young, multiple star system that probably is a Columba member, notwithstanding that its W component of -8.3$\pm3.7$ km s$^{-1}$ may be quite different from the mean W of Columba stars as estimated by Torres et al. (2008; see note to our Table 3).  On an A-star color-magnitude diagram (Zuckerman 2001) the A1-type primary is positioned in the vicinity of Pleiades stars.  A ROSAT HRI image presented by Schroder \& Schmitt (2007) indicates that the primary and its M-type companion $\sim$25" to the north are both strong X-ray sources; no doubt the primary has a close (spatially unresolved) companion of spectral type intermediate between it and the M-type star.  The optical secondary is probably of mid-M type, although it is difficult to deduce the exact M subclass because of apparent disagreement between comments in Schroder \& Schmitt (2007) and data in Vizier.  In any event, the absolute K mag (5.4)  of the M-star is such that on a color-mag diagram it lies well above old mid-M type stars (e.g., Figure 2 in Zuckerman \& Song 2004), and thus is consistent with the 30 Myr age of Columba.
\vskip 0.1in  
\noindent HIP 14551: this is a $\sim$70" binary star.  The M4.5 secondary (based on its I-K = 2.5) is located 70" to the SSW of the primary.  On an A-star color-mag diagram the A5 primary lies near the very young star $\beta$ Pictoris.  On a color-mag diagram the absolute K magnitude (6.3) of the secondary places it well above old, M4.5, Gliese stars (e.g., Figure 2 in Zuckerman \& Song 2004). 
\vskip 0.1in
\noindent HIP 14913:  Like HIP 12413 discussed above, HIP 14913 is a triple star system; The AB separation is about 0.7" and AC are separated by about 3.7".  C is a K-type star.  Concerning entries in Table 3, the lithium line EW in the F6 primary = 65 m\AA\ while f$_x$ =-4.06 for the entire triple system considered as a single star.  Our SSO spectra indicate that two nearby K-type stars CD-46 1064 (TYC 8060-1673-1) and CD-44 1173 (TYC 7574-803-1) are also Tuc/Hor members, and were so identified by Torres et al. (2008).   In addition, the Hipparcos measured parallax for HIP 14913 is in good agreement with the photometric parallax of the two K-type Tycho stars calculated independently by us and by Torres et al (2008).  
\vskip 0.1in
\noindent HIP 17248: The lithium line is too weak to be measured and f$_x$ = -3.36; both measurements are consistent with a 30 Myr old M0.5 star.  On a color-mag diagram its absolute K-mag (4.64) places HIP 17248 above the main sequence.  Thus, all age indicators and UVW are consistent with membership in the Columba Association although -- because of the negative value of W -- probably not with membership in Tuc/Hor. 
\vskip 0.1in
\noindent HIP 17782:  The star is an $\sim$equal brightness 0.36" binary.  The lithium and f$_x$ entries in Table 3 pertain to the two stars considered together.
\vskip 0.1in
\noindent HIP 17797: This is an 8" binary composed of an A1 and an A2 star.  The stars lie near and below the location of a typical Pleiades star on an A-star color-mag diagram.  The system is a weak X-ray source, suggesting the presence of a third star.
\vskip 0.1in
\noindent HIP 23179: This is a 5" binary composed of an A1 and a G0 star.  In the secondary the lithium line EW = 165 m\AA.  The system is a strong RASS X-ray source.  
\vskip 0.1in
\noindent HIP 23362:   The star lies near the location of a typical Pleiades star on an A-star color-mag diagram. 
\vskip 0.1in
\noindent HD 36869: This star is AH Lep.  The spectrum of the star has been measured previously by Cutispoto et al. (2002), Wichmann et al. (2003), and Lopez-Santiago et al (2010).
\vskip 0.1in
\noindent HIP 26309: On an A-star color-mag diagram the star lies near the very young star $\beta$ Pictoris.  
\vskip 0.1in
\noindent HIP 26990:  The spectrum of the star has been measured previously by Cutispoto et al. (2002), Wichmann et al. (2003), and Waite et al. (2005).
\vskip 0.1in
\noindent HIP 28474:  The lithium EW and f$_x$ are both somewhat too small for a typical Tuc/Hor or Columba star.  But, because of the essentially perfect agreement of the UVW of HIP 28474 with that of the mean Columba UVW given in Torres et al (2008) along with the very plausible space location of HIP 28474 with respect to other Columba stars, we deem the star to be a likely Columba member.   
\vskip 0.1in
\noindent HIP 32104: On an A-star color-mag diagram the star lies near the very young star HD 141569.  HIP 32104 is a RASS source, suggesting the presence of a second star.   
\vskip 0.1in
\noindent HIP 83494: On an A-star color-mag diagram the star lies near the very young star $\beta$ Pictoris.  In both R.A. and Decl. HIP 83494 lies far from any Tuc/Hor or Columba star proposed by either Zuckerman \& Song (2004) or Torres et al (2008).  
\vskip 0.1in 
\noindent HIP 84642:  Chauvin et al (2010) resolve the star as a 0.22" binary with delta K$_s$ = 2.5 mag.  As noted by Chauvin et al, probably because of the large flux contrast and small angular separation between the primary and secondary, HIP 84642 does not appear in the Hipparcos double star catalog.  They deem the secondary to be of spectral type $\sim$M5 and the system age to be $\sim$40 Myr.

The UVW of HIP 84642 is in only fair agreement with the mean UVW of the Tuc/Hor Association given by Zuckerman \& Song (2004) and by Torres et al (2008) and HIP 84642 is located in a sky position that contains no (other) members of Tuc/Hor listed in either of these review articles.  Therefore, we regard HIP 84642 as a possible rather than likely member of Tuc/Hor.
\vskip 0.1in
\noindent BD+44 3670: Guillout et al (2009) and P. Guillout (2010, private communication) present data for this star including a lithium line EW = 196 m\AA.  The photometric distance, radial velocity, and UVW in Table 3 are from P. Guillout (2010, private comm.).   
\vskip 0.1in
\noindent HIP 116805: The star lies near the location of a typical Pleiades star on an A-star color-mag diagram. 

\subsection{HR 8799}

HIP 114189 (= HR 8799) is known to be orbited by a multiple system of massive planets imaged by Marois et al. (2008, 2010) and by a massive dusty debris disk (Rhee et al 2007a, Su et al 2009).   In Table 3 we place the star in the Columba Association based on its Galactic space motion and other age indicators mentioned below.  If, as we suggest in Section 5.2, the nucleus of the Columba Association lies near 5h30m R. A. and is $\sim$50 pc from Earth, then currently HR 8799 is $\sim$70 pc from the nucleus.  A peculiar velocity of 2 km$^{-1}$ over a period of 30 Myr would produce this separation.  Based on their Bayesian statistical analysis, Doyon et al (2010) independently deduce that HR 8799 is a member of the Columba Association.  

Recently, Currie et al (2011) also deduce a young age for HR 8799; models of dynamical stability and planet evolution (cooling) lead them a preferred age near 30 Myr.  The near-IR colors of the planets of HR 8799 differ from those of conventional (old) L- and T-type substellar objects, but are similar to that of the $\sim$6 Jupiter mass 2M1207b (Marois et al 2008; Patience et al. 2010; Currie et al 2011).  As noted by Marois et al (2010), evolutionary models for cooling 30 Myr old planets suggest masses about 6 times that of Jupiter for the four known planets of HR 8799.  Given the well-established age of 2M1207b ($\sim$10 Myr), its unusual near-IR spectrum can most readily be attributed to atmospheric properties engendered by its low mass.    

HR 8799 lies below the position of a typical Pleiades star on an A-star color-mag diagram; see HR 8799 plotted on Fig. 5 of Zuckerman (2001).  Because HR 8799 is a $\lambda$ Boo type star with a peculiar surface composition, it probably should be plotted slightly to the right of where it appears in Zuckerman's Fig. 5, but still below the Pleiades line in the figure.  For example, with the T$_{eff}$ = 7430 K derived by Gray \& Kaye (1999), the star would plot near B-V = 0.31.  Relative to the locus of A-type Pleiades stars, HR 8799 plots as low or lower than all members of a sample of $\sim$20 $\lambda$ Boo stars with Hipparcos measured parallaxes identified by R. Gray (private comm. 2010).  Again this is consistent with a young age.

HR 8799 has one of the most massive dusty debris disks known for any main sequence star (Rhee et al. 2007a).  The range of dust temperatures is extensive (Su et al 2009); in their preferred model an inner warm dust belt with temperature $\sim$150 K extends between 6 and 15 AU from the star.  Marois et al (2010) note that the inner edge of such a dust belt could be in a 4:1 mean motion resonance with planet HR 8799e while the outer edge must be closer to the star than 15 AU to avoid planet e's chaotic region.  The location of the inner edge is determined only by a fit to the shorter wavelength portion of the Spitzer IRS spectrum combined with the assumption that the grains radiate like blackbodies at the relevant wavelengths (see Figures 3 and 4 in Su et al 2009). Because the radius of a typical grain must be at least a few microns to avoid radiative blowout, this assumption is valid.   Since the longer wavelength IRS emission is due to a blend of emission from the warm dust belt and cooler dust in an outer belt, the outer radius of the inner belt is not so well determined.  It remains to be seen whether yet a fifth planet can be squeezed in between the warm dust belt and planet e (Hinkley et al. 2011).

\subsection{Argus} 

Torres et al. (2008) proposed a new association comprising more than 60 stars of which somewhat more than half are members of the open cluster IC 2391, located $\sim$140 pc from Earth.   Torres et al dub the non-IC 2391 stars "field Argus members" and, of these, $\sim$half are more than 100 pc from Earth.  Indeed, only six of their field Argus stars are within the 70 pc radius sphere surrounding Earth that constrains our proposed Table 4 additions to Argus.  In addition, whereas none of the 29 Argus field members proposed by Torres et al are of spectral type earlier than F0, five of the Argus stars in Table 4 are A-type.  We therefore expect that many additional A-type Argus members will be identified beyond 66 pc from Earth.   Previously, Eggen (1991) proposed an IC 2391 Supercluster containing many early-type stars, including Table 2 stars HIP 98495 and HIP 57632 ($\beta$ Leo, only 11 pc from Earth).  In the future, some A-type members of Eggen's Supercluster more distant than 66 pc are likely to be joined to the Argus Association defined by Torres et al. (2008).

In addition to early-type Argus field stars more distant from Earth than $\sim$65 pc, IC 2391 contains some A- and B-type stars (e.g., Siegler et al. 2007).  Siegler et al. consider 34 members of IC 2391 ranging among spectral types B through M.  Only 12 stars in the Siegler sample are included among the 35 IC 2391 stars considered by Torres et al (2008) to be "high probability members" of the Argus Association.  Neither of these two papers cites the other which is understandable given that their submissions may have been nearly simultaneous.   However, the Siegler et al (2007) paper is also not cited in a 2009 paper by the Torres group (da Silva et al. 2009).  In any event, we consider the Argus Association and IC 2391 in Section 5.5.2 in conjunction with Table 11.

Torres et al (2008) suggest an age of 40 Myr for the Argus Association.  Its large negative U component of space motion distinguishes Argus from other young nearby moving groups. 
While Argus members listed by Torres et al (2008) are deep in the southern hemisphere, Table 2 stars HIP 57632 and HIP 99770 are in the northern hemisphere. 

The A-type stars HIP 50191, 57632 and 99770 all lie near the typical Pleiades star on an A-star color-mag diagram, while HIP 79797 and 98595 lie substantially below Pleiades stars.  The F-type star HIP 68994 also lies well below Pleiades stars; for it,  f$_x$ = -5.1.

\subsection{Spitzer Observations} 

Unveiling the evolution of dusty debris disks as a function of stellar age has been a major focus of Spitzer studies of main sequence stars.  Rebull et al. (2008) and Gaspar et al. (2009) each present tables listing stellar associations and clusters of known age and the fraction of such stars with Spitzer detected IR excess emission.  Neither listing contains stars in the AB Dor Association although a few such stars appear in Spitzer papers by Plavchen et al. (2009), Carpenter et al. (2009), and Hillenbrand et al. (2008).  The Rebull et al paper does include an entry for Tuc/Hor based on a few stars from her paper and from Smith et al. (2006); but many more Tuc/Hor stars are included in tables in the present paper.

\subsubsection{Individual spectral energy distributions (SED)}

\ \ \ \ \ \ \ \ \ \ \


Figures 1 - 7 display the SED for all stars in Tables 8-10 with definite or probable IR excess emission.  MIPS fluxes presented in this paper are color-corrected.  The tables give stellar and dust parameters derived from our SED fitting routine.  When only an upper limit to the 70 $\mu$m flux density is plotted, then an indicated dust temperature is a lower limit and the indicated dust luminosity is an upper limit.  We consider here a few of the more difficult and interesting SEDs.
\vskip 0.1in
\noindent CD-60 416 (Table 9):  The 24 $\mu$m MIPS flux density and the IRS spectrum both lie slightly above the estimated photospheric flux.  However, the IRS is noisy and is not rising toward long wavelengths, so we regard the apparent excess emission at 24 $\mu$m as questionable.
\vskip 0.1in
\noindent HIP 16563 (Table 8, Fig. 1):  This is $\sim$10" binary.  The M0 secondary appears to have 24 $\mu$m emission that is 40\% in excess of the photosphere.  If so, then HIP 16563B is one of only a handful of M-type stars with measurable excess IR emission.  The primary is about a magnitude brighter than the secondary and appears to have a 24 $\mu$m excess of about 30\%.   Lestrade et al (2006) found no evidence for cold dust at these stars.  Because the binary nature of the star introduces additional complexities into the analysis, we regard as tentative the excesses at both stars.   
\vskip 0.1in				
\noindent HIP 30034 (Table 9, Fig. 5):  This is AB Pic, a star that has a companion, imaged with AO, with mass ($\sim$13.5 Jupiter masses) that straddles the planet/brown dwarf boundary (Chauvin et al 2005).  Although the 8 $\mu$m flux density measured with IRAC on Spitzer is on or very near the photosphere, there is evidence for excess emission in both the 12 $\mu$m IRAS and 24 $\mu$m MIPS channels.  The MIPS excess is $\sim$25\%.  The color-corrected IRAS 12 $\mu$m flux density is 70 $\pm$11 mJy while the photosphere is $\sim$40 mJy. 
\vskip 0.1in		
\noindent HIP 32435 (Table 9, Fig. 5):   The shape of the IRS spectrum seems difficult to reconcile with the elevated 60 $\mu$m IRAS point. 						
\vskip 0.1in
\noindent HIP 68994 (Table 10, Fig. 7):  This star lies precisely in the Galactic plane so that contamination by background IR sources is always a possibility.  That said, we note the elevated AKARI  9 $\mu$m flux density that, along with the 24 $\mu$m MIPS measurement, suggests the presence of substantial quantities of warm dust particles. 
\vskip 0.1in
\noindent HIP 114189 (Note to Table 9): The SED has been analyzed in great detail by Su et al (2009), thus we do not present it here.
\vskip 0.1in
\noindent HIP 115738 (Table 8, Fig. 2):  The IRS spectrum of this A0 star indicates a rising SED from 10 to 30 $\mu$m, but with the MIPS 70 $\mu$m flux density essentially back on the photosphere.  We have not tried to fit a dust temperature.

\subsubsection{Evolution of dusty debris disks with time} 

\ \ \ \ \ \ \ \ \

As mentioned in Section 4, because the Columba Association is about the same age as Tuc/Hor (30 Myr) we combine Spitzer observed stars from these associations into a single entry in Table 11.  Both Rebull et al (2008) and Gaspar et al (2009) tabulate the percentage of stars with excess emission found by Spitzer in various nearby clusters and associations.  For the five associations/clusters for which we agree with the fraction of members with excess 24 and 70 $\mu$m emission given in Table 5 in Rebull et al (2008) we cite their paper in the right hand column of Table 11 and adopt their quoted values; and similarly for the three clusters where we cite Gaspar.  For the 24 $\mu$m excess fraction in IC 2391 we follow both Gaspar and Rebull (who agree).  For the Pleiades and for NGC 2451 we take excess fractions from the original (cited) Spitzer papers.

At 24 $\mu$m wavelength Spitzer was sufficiently sensitive to detect the stellar photospheres.  The entries in the third and fourth columns of Table 11 indicate that the fraction of stars that possess excess 24 $\mu$m emission above the photosphere is rather constant at about 1/3 between 8 and $\sim$50 Myr.  Subsequently, this fraction declines to $\sim$15\% at $\sim$100 Myr, and then to only a percent or two at $\sim$700 Myr.

At 70 $\mu$m, because the photospheric flux level is usually not reached in Spitzer observations, only lower limits to the percentage of stars with excess emission can be derived.  This lower limit is about 1/3 for ages between 8 and 50 Myr, i.e., comparable to the percentage of stars with excess 24 $\mu$m at similar ages.  For cluster stars of age 100 Myr and greater the fraction that possess 70 $\mu$m excess emission is much smaller than 1/3. 

\section{Conclusions}

We propose 35 star systems within $\sim$70 pc of Earth as new members of previously identified young, moving groups.  With but one exception, these 35 stars are brighter than 10th magnitude at V-band.  Thus, they should generally be excellent targets for extreme-AO and space-based, near-infrared, imaging searches for warm planets.

Among the 35 star systems are some that appear to have the same Galactic space motions and ages as stars in the previously proposed Columba and Argus field associations (Torres et al 2008), but over a much wider range of right ascension and/or declination.  It remains to be seen whether all such stars belong to these already defined kinematic groups or if additional young moving groups will need to be defined.  Our unpublished optical spectroscopic observations from Siding Spring and Lick observatories of X-ray bright stars indicate that many young stars near Earth have space motions that differ from those of all moving groups and associations listed in Zuckerman \& Song (2004) and Torres et al (2008).  

We comment on many of the 35 star systems and also on some other stars that clearly are young but that do not seem to quite fit into known moving groups.   One of the more interesting of the 35 stars is HR 8799 that is orbited by at least four giant planets at wide separations.  The Galactic space motion of HR 8799 is in excellent agreement with that of the 30 Myr old Columba Association and we present it as one of six northern hemisphere stars we are proposing as members of this association.

We present the first comprehensive consideration of Spitzer data for stars in the AB Doradus and Tucana/Horologium Associations.  We also consider Spitzer results for nearby stars in the Columba and field Argus Associations.   As young stars are wont to do, many of these stars display excess IR emission at 24 and/or 70 $\mu$m wavelength.  For a few stars, 24 $\mu$m emission appears to dominate thus suggesting the presence of warm dust particles.  One such warm dust, solar-like, star may be AB Pic which is known to be orbited by a distant companion with mass that straddles the planet/brown dwarf mass boundary.

Combination of our Spitzer results with those for 11 other nearby clusters and associations illustrates the decay of dusty debris with time over the age range 8 to 750 Myr.  For cluster/association stars of ages 8 to $\sim$50 Myr about 1/3 display excess 24 and 70 $\mu$m emission above the photosphere.  The percentage with excess 24 $\mu$m emission drops to $\sim$15\% at 100 Myr and then to only a few percent at $\sim$700 Myr.  Similarly, the percentage of stars with Spitzer detected excess emission at 70 $\mu$m and age $>$100 Myr is much smaller than 1/3.
 
\vskip 0.2in

We thank the referee for useful suggestions.  This research was funded in part by NASA grants to UCLA and the University of Georgia.

\section*{References}
\begin{harvard}

\item[Aumann, H. et al. 1984, ApJ 278, L23]
\item[Balog, Z. et al. 2009, ApJ 698, 1989]
\item[Carpenter, J. et al. 2009, ApJS, 181, 197]
\item[Chauvin, G. et al. 2005, A\&A 438, L29]
\item[Chauvin, G. et al. 2010, A\&A 509, A52]
\item[Currie, T. et al. 2011, ApJ 729,128]
\item[Cutispoto, G., Pastori, L., Pasquini, L., de Medeiros, J., Tagliaferri, G. \& Andersen, J. 2002, ] 
\indent A\&A 384, 491
\item[da Silva, L., Torres, C., de La Reza, R., Quast, G., Melo, C. \& Sterzik, M. 2009, A\&A 508, 833]
\item[Doyon, R., Malo, L., Lafreniere, D. \& Artigau, E. in Proceedings of the conference "In the Spirit of]
\indent Bernard Lyot": Oct. 25-29,  2010. LESIA/CNRS, Paris Observatory.
\item[Eggen, O. J. 1991, AJ 102, 2028]
\item[Engelbracht, C. et al. 2007, PASP 119, 994]
\item[Fernandez, D., Figueras, F. \& Torra, J. 2008, A\&A 480, 735]
\item[Gaspar, A. et al. 2009, ApJ 697, 1578]
\item[Gizis, J. 2002, ApJ 575, 484]
\item[Gontcharov, G. 2006, Astron. Lett. 32, 759] 
\item[Gorlova, N., Rieke, G., Muzerolle, J., Stauffer, J., Siegler, N., Young, E. \&  Stansberry, J. 2006, ApJ]
\indent 651, 1130 
\item[Gray, R. \& Kaye, A. 1999, AJ 118, 2993]
\item[Guillout, P. et al. 2009, A\&A 504, 829]
\item[Habing, H. et al. 2001, A\&A 365, 545]
\item[Hauschildt, P., Allard, F. \& Baron, E. 1999, ApJ 512, 377]
\item[Hillenbrand, L. et al. 2008, ApJ 677, 630]
\item[Hinkley, S. Carpenter, J., Ireland, M. \& Kraus, A. 2011, ApJL 730, L21]
\item[Kastner, J. H., Zuckerman, B., Weintraub, D. \& Forveille, T. 1997, Science 277, 67]
\item[Kiss, L. et al. 2011, MNRAS 411, 117]
\item[Lestrade, J.-F., Wyatt, M., Bertoldi, F., Dent, W. \&  Menten, K. 2006, A\&A 460, 733]
\item[Looper, D., Bochanski, J., Burgasser, A., Mohanty, S., Mamajek, E., Faherty, J.,]
\indent West, A. \& Pitts, M. 2010b, AJ 140, 1486L
\item[Looper, D., Burgasser, A., Kirkpatrick, J. D. \& Swift, B. 2007, ApJ 669, L97]
\item[Looper, D. et al. 2010a, ApJ 714, 45L]
\item[Lopez-Santiago, J., Montes, D., Crespo-Chacon,. I. \& Fernandez-Figueroa, M. 2006, ApJ 643, 1160]
\item[Lopez-Santiago, J. et al. 2010, A\&A 514, A97]
\item[Mamajek, E. 2005, ApJ 634, 1385]
\item[Marois, C., Macintosh, B., Barman, T., Zuckerman, B., Song, I., Patience, J., Lafreniere, D. \& ] 
\indent Doyon, R. 2008, Science 322, 1348
\item[Marois, C., Zuckerman, B., Konopacky, Q., Macintosh, B. \& Barman, T. 2010, Nature 468, 1080] 
\item[Melis, C., Zuckerman, B., Rhee, J. \& Song, I. 2010, ApJ 717, L57]
\item[Moor, A., Abraham, P., Derekas, A., Kiss, Cs., Kiss, L., Apai, D., Grady, C. \& Henning, Th. ]
\indent 2006, ApJ 644, 525
\item[Moor, A. et al. 2009, ApJ 700, L25]
\item[Patience, J., King, R., de Rosa, R. \& Marois, C. 2010, A\&A 517, A76]
\item[Plavchan, P., Werner, M., Chen, C., Stapelfeldt, K., Su, K, Stauffer,J. \& Song, I. 2009, ApJ ] 
\indent 698, 1068
\item[Rebull, L. et al. 2008, ApJ 681, 1484]
\item[Rhee, J., Song, I. \& Zuckerman, B 2007b, ApJ 671, 616]
\item[Rhee, J., Song, I., Zuckerman, B. \& McElwain, M. 2007a, ApJ 660, 1556]
\item[Rodriguez, D., Bessell, M., Zuckerman, B. \& Kastner, J.H. 2011, ApJ 727, 62R]
\item[Schlieder, J., Lepine, S. \& Simon, M. 2010, AJ 140, 119]
\item[Schroder, C. \& Schmitt, J. 2007, A\&A 475, 677]
\item[Shkolnik, E., Liu, M.,  Reid, I. N., Dupuy, T. \& Weinberger, A. 2011, ApJ 727, 6S]
\item[Siegler, N., Muzerolle, J., Young, E., Rieke, G., Mamajek, E., Trilling, D., Gorlova, N. \& Su, ] 
\indent K. 2007, ApJ 654, 580
\item[Silverstone, M. 2000, Ph.D. thesis, Univ. Calif. Los Angeles, 194 pp.]
\item[Scholz, R.-D., McCaughrean, M., Zinnecker, H. \& Lodieu, N. 2005, A\&A 430, L49]
\item[Smith, P., Hines, D., Low, F., Gehrz, R., Polomski, E. \& Woodward, C. 2006, ApJ 644, 125]
\item[Song, I., Zuckerman, B. \& Bessell, M. 2003, ApJ 599, 342; Erratum 2004, ApJ 603, 804]
\item[Su, K. et al. 2009, ApJ 705, 314]
\item[Torres, C., da Silva, L., Quast, G., de la Reza, R. \& Jilinski, E. 2000, AJ 120, 1410]
\item[Torres, C., Quast, G., da Silva, L., de la Reza, R., Melo, C. \& Sterzik, M. 2006, A\&A 460, 695]
\item[Torres, C., Quast, G., Melo, C. \& Sterzik, M. 2008, "Handbook of Star Forming Regions,] \indent Volume II: The Southern Sky" ASP Monograph Publ., Vol. 5., ed. Bo Reipurth, p.757
\item[Trilling, D. et al. 2008, ApJ 674, 1086]
\item[Voges, W. et al. 1999, A\&A 349, 389]
\item[Voges, W. et al. 2000, IAU Circ. \#7432, 3]
\item[Waite, I., Carter, B., Marsden, S. \& Mengel, M. 2005, PASA 22, 29] 
\item[Webb, R., Zuckerman, B., Platais, I., Patience, J., White, R. J., Schwartz, M., \& McCarthy, C.] \indent 1999, ApJ 512, L63
\item[Wichmann, R., Schmitt, J.H.M.M. \& Hubrig, S. 2003, A\&A 399, 983] 
\item[Zuckerman, B. 2001, ARAA 39, 549]
\item[Zuckerman, B., Bessell, M. S., Song, I. \& Kim, S. 2006, ApJ 649, L115]
\item[Zuckerman, B. \& Song, I. 2004, ARAA 42, 685]
\item[Zuckerman, B., Song, I. \& Bessell, M. 2004, ApJ 613, L65]
\item[Zuckerman, B., Song, I., Bessell, M. \& Webb, R. 2001a, ApJ 562, L87]
\item[Zuckerman, B., Song, I. \& Webb, R. 2001b, ApJ 559, 388]
\item[Zuckerman, B. \& Webb, R. 2000, ApJ 535, 959]
\item[Zuckerman, B., Webb, R., Schwartz, M. \& Becklin, E. E. 2001c, ApJ 549, L233]

\end{harvard}

\clearpage
\begin{table}
\caption\noindent{Papers Proposing Memberships in the Nearest Known Young Stellar
Associations}
\begin{tabular}{@{}lcccc}
\br
Paper& Tuc/Hor& AB Dor& TW Hya& $\beta$ Pic\\
\mr
Kastner et al. 1997& & & X& \\
Webb et al. 1999& & & X& \\
Zuckerman \& Webb 2000& X& & & \\
Torres et al. 2000& X& & & \\
Zuckerman et al. 2001a& & & & X \\
Zuckerman et al. 2001b& X& & & \\
Zuckerman et al. 2001c& & & X& \\
Gizis 2002& & & X& \\
Song et al. 2003& X& & X& X \\
Zuckerman et al. 2004& & X& & \\
Zuckerman \& Song 2004& X& X& X& X\\
Scholz et al. 2005& & & X& \\
Mamajek 2005& & & X& \\
Torres et al. 2006& & & & X\\
Lopez-Santiago et al. 2006& & X& & \\
Looper et al. 2007& & & X& \\
Torres et al. 2008& X& X& X& X \\
Fernandez et al. 2008& X& X& X& X \\
da Silva et al. 2009& X& X& X& X \\
Schlieder et al. 2010& & X& & X \\
Looper et al. 2010a \& b& & & X& \\
Rodriguez et al. 2011& & & X& \\
Kiss et al. 2011& X& & & X\\
Shkolnik et al. 2011& & & X& \\
This paper& X& X& &\\
\br
\end{tabular}
\end{table}
\noindent Note $-$ In addition to the above major young stellar associations nearest to Earth, Zuckerman et al. (2006) proposed the somewhat older and sparser, but comparably nearby, "Carina-Near" moving group.  The Carina Association proposed by Torres
et al (2008) has essentially nothing in common with the Carina-Near group; the latter is
much nearer to Earth and much older than the former and
has a much more negative U component of space motion.  While the Carina, Columba and Argus Associations possess some members close to Earth (see, e.g., Tables 3 \& 4), as defined by Torres et al (2008), stars in these three associations are, on average, substantially more distant from Earth than are stars in the four Table 1 Associations (see Table 2 in Torres et al.).

\clearpage

\begin{table}
\caption{Proposed AB Doradus Moving Group Members}
\begin{tabular}{@{}lccccccccc}
\br
HIP& HD& R.A.& Decl.& Spec.& V& Dist.& Rad. Vel.& (U,V,W)& UVW error\\
& & (h/m)& (deg)& Type& (mag)& (pc)& (km s$^{-1}$)& (km s$^{-1}$)& (km s$^{-1}$)\\
\mr
13209& 17573& 02 49& +27& B8& 3.6& 49& 4.0$\pm$4.1& -8.5,-25.7,-15.6& 3.2,1.9,2.0\\
15353& 20888& 03 17& -66& A3& 6.0& 58& 26.0$\pm$0.5& -6.8,-27.5,-11.5& 0.4,0.5,0.4\\
93580& 177178& 19 03& +01& A4& 5.8& 55& -23.1$\pm$2.3& -11.3,-24.3,-12.8& 1.9,1.4,0.6\\
95347& 181869& 19 23& -40& B8& 4.0& 52& -0.7$\pm$4.1& -8.1,-26.0,-14.6& 3.8,1.4,1.8\\
109268& 209952& 22 08& -46& B6& 1.7& 31& 10.9$\pm$1.7& -7.0,-25.6,-15.5& 1.1,0.7,1.4\\
115738& 220825& 23 26& +01& A0& 4.9& 50& -4.4$\pm$0.6& -7.0,-26.4,-13.3& 0.3,1.0,0.8\\
117452& 223352& 23 48& -28& A0& 4.6& 44& 8.7$\pm$2.0& -7.0,-27.5,-13.3& 0.6,1.4,2.0\\
\br
\end{tabular}
\end{table}
\noindent Notes $-$ Input data (R.A., Decl., distance, proper motion) for the UVW calculations in Tables 2-4 are from the Hipparcos catalog.  UVW are defined with respect to the Sun, with U positive toward the Galactic Center, V positive in the direction of Galactic rotation, and W positive toward the North Galactic pole.  Characteristic mean UVW for AB Dor group stars are given in Zuckerman \& Song (2004) as -8, -27, -14 km s$^{-1}$, and in Torres et al. (2008) as -6.8$\pm$1.3, -27.2$\pm$1.2, -13.3$\pm$1.6 km s$^{-1}$. 
HIP 93580 and 117452 are considered in Section 5.1.

\clearpage

\begin{landscape}
\begin{table}
\caption{Proposed Tucana/Horologium \& Columba Association Members}
\begin{tabular}{@{}lccccccccccc}
\br
HIP& HD& R.A.& Decl.& Spec.& V& Dist.& Rad. Vel.& (U,V,W)& UVW error& Li EW& f$_x$\\
& & (h/m)& (deg)& Type& (mag)& (pc)& (km s$^{-1}$)& (km s$^{-1}$)& (km s$^{-1}$)& (m\AA)& \\
\br
1134& 984& 00 14& -07& F5*& 7.3& 46& -2.2$\pm$1.2& -12.2,-23.2,-6.0& 0.7,1.3,1.2& 120& -4.26 \\
12413& 16754& 02 39& -42& A1*& 4.7& 40& 18.0$\pm$4.2& -10.8,-21.2,-8.3& 0.6,1.9,3.7& & \\ 
12925& 17250& 02 46& +05& F8& 7.9& 63& 4.3$\pm$1.1& -10.8,-23.6,-0.6& 1.0,1.8,0.9& 145& -4.33 \\ 
14551& 19545& 03 07& -27& A5& 6.2& 58& 13.8$\pm$0.8& -10.8,-20.4,-3.2& 0.4,0.8,0.8\\
14913& 20121& 03 12& -44& F6& 5.9& 44& 13.5$\pm$2.1& -10.8,-18.8,-2.3& 0.5,1.2,1.8& 65& -4.06 \\
17248& & 03 41& +55& M0.5*& 11.2& 37& -3.2$\pm$0.6& -12.0,-23.5,-6.5& 1.2,1.7,0.6& & -3.36 \\
17764& 24636& 03 48& -74& F3& 7.1& 55& 15.5$\pm$1.3& -8.6,-21.8,-2.5& 0.6,1.0,0.8& 60& -5.4 \\
17782& 23524& 03 48& +52& G8& 8.8& 51& -2.2$\pm$0.6& -9.5,-20.1, -4.5& 1.8,2.9,1.0& 243& -3.16 \\ 
17797& 24071J& 03 48& -37& A1& 4.7& 49& 15.6$\pm$0.4& -10.2,-21.5,-1.0& 0.3,0.6,0.6 \\
23179& 31647& 04 59& +37& A1*& 5.0& 49& 7.7$\pm$2.5& -12.6,-22.2,-5.8& 2.5,1.2,0.3 \\
23362& 32309& 05 01& -20& B9*& 4.9& 64& 24.2$\pm$2.8& -13.8,-22.6,-5.5& 1.8,1.6,1.5\\
& 36869& 05 34& -15& G3*& 8.5& (59)& 23.0$\pm$1.0& -12, -21,-5&  & 204&  -3.53 \\
26309& 37286& 05 36& -28& A2*& 6.3& 57& 22.4$\pm$1.2& -11.3,-20.0,-4.9& 0.7,0.9,0.6\\
26453& 37484& 05 37& -28& F3*& 7.2& 60& 23.5$\pm$0.4& -11.9,-20.9,-5.4& 0.3,0.4,0.4& 87& -5.0 \\
26990& 38397& 05 43& -39& G0*& 8.1& 52& 22.8$\pm$0.6& -11.6,-20.3,-5.1& 0.3,0.5,0.4& 137& -4.05 \\
28474& 41071& 06 00& -44& G8*& 9.1& 54& 23.8$\pm$0.4& -12.1,-21.1,-5.7& 0.3,0.4,0.5& 155& -4.41 \\
32104& 48097& 06 42& +17& A2*& 5.2& 43& 15.0$\pm$4.2& -10.4,-20.1,-4.9& 4.0,2.0,0.9\\
83494& 154431& 17 03& +34& A5& 6.1& 54& -21.5$\pm$1.4& -10.0,-24.3,-0.2& 0.7,1.1,1.0\\
84642& 155915& 17 18& -60& G8& 9.5& 55& 1.3$\pm$0.7& -12.6,-24.8,-1.1& 1.2,1.9,0.3\\
& **BD+44& 21 00& +45& G2*& 8.8& (65)&  -23.2$\pm$1.5&-11.0,-22.8,-8.2&  & 196& -3.48 \\
114189& 218396& 23 07& +21& A5*& 6.0& 40& -12.6$\pm$1.4& -12.3,-21.5,-7.2& 0.5,1.2,1.0\\
116805& 222439& 23 40& +44& B9*& 4.1& 52& -12.7$\pm0.6$& -11.7,-20.3,-5.9& 0.6,0.8,0.4\\
\br
\end{tabular}
\end{table}
\end{landscape}
\noindent Notes $-$ **BD+44 = BD+44 3670.  In the 5th column, a * after a spectral type indicates a suggested Columba Association member.  The listed distances to HD 36869 and BD+44 3670 are derived photometrically.  Tycho-2 proper motions are used in the calculation of UVW for these two stars.  The UVW of HIP 14551 and 17782 appear to be a mixture of those of Columba and Tuc/Hor.   Characteristic mean UVW for Tuc/Hor stars are given in Zuckerman \& Song (2004) as -11, -21, 0 km s$^{-1}$, and in Torres et al. (2008) as -9.9$\pm$1.5, -20.9$\pm$0.8, -1.4$\pm$0.9 km s$^{-1}$.  Torres et al give a mean UVW for Columba stars of -13.2$\pm$1.3, -21.8$\pm$0.8, -5.9$\pm$1.2 km s$^{-1}$.  See Sections 4 and 5.2 for additional details regarding Table 3 stars.  HIP 114189 (= HR 8799, see Section 5.3) is orbited by at least four giant planets. 

\clearpage

\begin{table}
\caption{Proposed Argus Association Members}
\begin{tabular}{@{}lccccccccc}
\br
HIP& HD& R.A.& Decl.& Spec.& V& Dist.& Rad. Vel.& (U,V,W)& UVW error\\
& & (h/m)& (deg)& Type& (mag)& (pc)& (km s$^{-1}$)& (km s$^{-1}$)& (km s$^{-1}$)\\
\mr
50191& 88955& 10 14& -42& A2& 3.8& 31&  7.4$\pm$2.7& -22.0,-10.5,-4.9& 0.5,2.6,0.6\\
57632& 102647& 11 49& +14& A3& 2.1& 11& -0.2$\pm$0.5&-20.1,-16.2,-7.6& 0.2,0.2,0.5\\
68994& 123058& 14 07& -61& F4& 7.8& 66& -5.2$\pm$1.0& -21.2,-12.1,-2.5& 1.4,1.4,0.3\\
79797& 145689& 16 17& -67& A4& 6.0& 55& -9.0$\pm$4.3& -23.2,-12.0,-4.9& 3.3,2.7,1.0\\
98495& 188228& 20 00& -72& A0& 4.0& 32& -6.7$\pm$0.7&  -21.8,-10.8,-4.5&  0.6,0.5,0.4\\
99770& 192640& 20 14& +36& A2& 5.0& 41& -17.3$\pm$2.8& -22.5,-11.7,-3.9& 0.8,2.7,0.1\\
 \br
\end{tabular}
\end{table}
\noindent Notes $-$ Additional information regarding the listed stars can be found in Section 5.4. Torres et al (2008) give a mean UVW for Argus stars of -22.0$\pm$0.3, -14.4$\pm$1.3, -5.0$\pm$1.3 km s$^{-1}$.    

\clearpage

\def\mc{\multicolumn}
\def\noIR{\tiny no IR source}
\def\str{\tiny Strange spec.}
\begin{landscape}
\begin{longtable}{lcccccccccc}
\caption{AB Doradus Stars Observed by Spitzer}\\
\hline
HIP& HD& name& spect& dist& ZS04& T08& Spitzer& AORKEY& Date of& Integ. time (sec)\\
& & & type& (pc)&  &  & Program ID& Archive ID& Observation& 24$\mu$m \ 70$\mu$m \\
\hline
\hline
\endfirsthead
\hline
\caption[]{(continued)}\\
HIP& HD& name& spect& dist& ZS04& T08& Spitzer& AORKEY& Date of& Integ. time (sec)\\
& & & type& (pc)&  &  & Program ID& Archive ID& Observation& 24$\mu$m \ 70$\mu$m \\
\hline
\hline
\hline
\endhead
\hline
\hline
\multicolumn{5}{r}{\small\sl continued on next page}\\
 \hline
 \endfoot
 \hline
 \endlastfoot
&        1405&    PW And&          K2&      (28)&            Y&       Y&          3600&            11254784&        2004-12-25&      48 \ \ \ \ \    252\\
3589&    4277&  &                  F8&      49&              Y&       Y&        3600&            11255040&        2005-01-28&      48 \ \ \ \ \  252\\
5191&    6569&   &                 K1&      50&              Y& &        3600&            11255296&        2004-12-25&      48  \ \ \ \ \       378 \\
6276&   &         BD-12 243&       G9&      35&              Y&       Y&        148&             5346304&         2004-07-11&      96 \ \ \ \ \     252\\
10272&   13482&   &                K1&      32&              Y&       Y&       3600&            11255808&        2005-01-29&      48 \ \ \ \ \    252\\
13027&   17332&   &                G0&      33&              Y&       Y &       3600&            11256832&        2005-01-26&      48\ \ \ \ \      252\\
14684&   19668&   IS Eri&          G0&      40&   &                   Y &          148&             5340928&         2005-01-24&      96\ \ \ \ \      252\\
14807&    &       BD+21 418B&      K6&      50&              Y&       Y&       72&              4541952&         2004-02-24&      332 \ \ \ \    630\\
14809&   &        BD+21 418&       G5&      50&              Y&       Y &       72&              4541952&         2004-02-24&      332 \ \ \ \     630\\
15353&  20888&  & A3& 58& &  &     10&              3697408&         2004-11-07&      48\ \ \ \ \      126\\
16563&   21845&   V577 Per&        G5&      34&              Y&       Y&       3600&            11257600&        2005-02-01&      48 \ \ \ \ \     252\\
&        21845B&  &                M0&      34&              Y&       Y&  &  &  & \\
17695&  &         G80-21&          M3&      16&              Y&       Y &       3600&            11258368&        2004-09-23&      48 \ \ \ \ \     378\\
18859&   25457&   HR 1249&         F5&      19&              Y&       Y &       148&             5308672&         2004-09-20&      96 \ \ \ \ \     252\\
19183&   25953&   &                F5&      55&              Y&       Y &       20707&           15009792&        2006-02-21&      96  \ \ \ \ \     252\\
25283&   35650&   &                K7&      18&              Y& &       3600&            11260416&        2005-03-07&      48 \ \ \ \ \      252 \\
25647&   36705A&  AB Dor&          K0&      15&              Y&       Y&          80&              4638720&         2004-02-21&      48\ \ \ \ \      252\\
&  36705B&  AB DorB&         M4&      15&              Y&       Y&  & & & \\
26369&    &  &                       K7&      24&              Y&       Y&       148&             6599680&         2004-09-21&      96\ \ \ \ \      252\\
26373&   37572&   UY Pic&          K0&      24&              Y&       Y &       148&             6599680&         2004-09-21&      96\ \ \ \ \      252\\
&  &  CD-35 2722&      M1&      (24)&     &               Y&      731&             26807552&        2008-11-25&      664 \ \ \ \     1260\\
30314&   45270&     &              G1&      23&              Y&       Y&       148&             6599424&         2004-06-22&      96\ \ \ \ \      252\\
&  &               GSC 08894-0426&  M3&      (22)&            Y&       Y&   3600&            11262464&        2004-12-25&      48\ \ \ \ \      378\\
31711&   48189&   AK Pic&          G2&      22&              Y&       Y&       80&              4639232&         2204-02-21&      48\ \ \ \ \      252\\
31878&  &         CD-61 1439&      K7&      22&              Y&       Y&       3600&            11262720&        2004-11-08&      48\ \ \ \ \      252\\
&  &                BD+20 1790&      K5&      (26)&   &                 Y &      148&             5348608&         2004-10-14&      96\ \ \ \ \      1260\\
36349&  &           V372 Pup&        M2&      15&              Y&       Y&       3600&            11264000&        2004-11-07&      48\ \ \ \ \      252\\
51317&  &         GJ 393&          M2&      7&     &                  Y&      40454&           22014720&        2009-01-02&      48\ \ \ \ \      756\\\
63742&   113449&  PX Vir&          K1&      22&              Y&       Y&       80&              4627968&         2005-06-22&      48\ \ \ \ \      252\\
76768&   139751&  &                  K5&      42&              Y&       Y &       3600&            11265536&        2005-08-26&      48\ \ \ \ \      252\\
81084&   &        NLTT 43056&      M0&      32&              Y&       Y &       3600&            11265792&        2005-03-10&      48\ \ \ \ \      378\\
82688&   152555&  &                G0&      48&              Y&       Y &      148&             5330944&         2005-03-10&      96\ \ \ \ \      252\\
&        317617&    &              K3&      (56)&     &               Y&       30594&           20485632&        2006-10-07& 30\ \ \ \ \    15  \\
86346&   160934&  GJ 4020A&        K7&      28&              Y&       Y&       72&              4554240&         2004-01-29&      192\ \ \ \     630\\
93580& 177178&  & A4& 55&  &  &   10&              3724544&         2004-04-11&      48\ \ \ \ \      126\\
106231&  &        LO Peg&          K7&      25&              Y&       Y&      80&              4641024&         2004-06-22&      48\ \ \ \ \      630\\
107948&  &          GJ 4231&         M2&      30&   &                   Y&      50356&          26063616&        2008-06-27&      48\ \ \ \ \      252\\
109268&  209952&  &  B6& 31&  &  &  713&             7345152&         2003-11-03&      96\ \ \ \ \      114\\
110526&  &          GJ 856&          M3&      16&              Y&       Y&      3600&            11266560&        2004-12-02&      48\ \ \ \ \      378\\
113579&  217343&  &                G4&      32&              Y&       Y&      148&             5269760&         2004-06-21&      96\ \ \ \ \      252\\
113597&  217379&  &                K8&      31&              Y&       Y&      123&             5022976&         2004-11-02&      144\ \ \ \     630\\
114066&  &        GJ 9809&         M1&      25&              Y&       Y&      3600&            11266816&        2004-10-15&      48\ \ \ \ \      378\\
114530&  218860&  &                G5&      51&              Y&       Y&      3600&            11267072&        2004-10-19&      48\ \ \ \ \     252\\
115738& 220825&  &  A0& 50&  & &     171&       3731456&         2004-07-11&      48\ \ \ \ \      126\\
117452&  220825&  & A0& 44&  &  &  173&             3731968&         2004-12-02&      48\ \ \ \ \      126\\
118008&  224228&  GJ 4377&         K2&      22&              Y&       Y&      3600&            11267584&        2004-11-04&      48\ \ \ \ \      252\\
\hline
\end{longtable}
\noindent Note $-$  Distances in parenthesis are estimated photometrically.   ZS04 and T08 refer to AB Dor members listed in Zuckerman \& Song (2004) and Torres et al (2008), respectively.
\end{landscape}

\clearpage

\def\mc{\multicolumn}
\def\noIR{\tiny no IR source}
\def\str{\tiny Strange spec.}
\begin{landscape}
\begin{longtable}{lcccccccccc}
\caption{Tucana/Horologium \& Columba Stars Observed by Spitzer}\\
\hline
HIP& HD& name& spect& dist& ZS04& T08& Spitzer& AORKEY& Date of& Integ. time (sec)\\
& & & type& (pc)&  &  & Program ID& Archive ID& Observation& 24$\mu$m \ 70$\mu$m \\
\hline
\hline
\endfirsthead
\hline
\caption[]{(continued)}\\
HIP& HD& name& spect& dist& ZS04& T08& Spitzer& AORKEY& Date of& Integ. time (sec) \\
& & & type& (pc)&  &  & Program ID& Archive ID& Observation& 24$\mu$m \ 70$\mu$m \\
\hline
\hline
\hline
\endhead
\hline
\hline
\multicolumn{5}{r}{\small\sl continued on next page}\\
 \hline
 \endfoot
 \hline
 \endlastfoot
490& 105& & G0& 40& Y& Y &         148    &         5295872  &       2003-12-10 &     96\ \ \ \ \     252\\
1113& 987& & G7& 44& Y& Y&        102       &      9022976&         2004-11-05    &  48\ \ \ \ \      126\\
1481& 1466&  & F8& 41& Y& Y&        72&              4539648  &       2004-12-02&      96\ \ \ \ \      630\\
1910&  & &           M1&      45&      Y&       Y&        72   &           4539904&         2004-06-22 &     332\ \ \ \      630\\
1993&   &         CT Tuc&          M0&      40&        Y&       Y &        72&              4540160   &      2004-06-22&      332\ \ \ \  630\\
2484&    2884&    beta1 Tuc&       B9&      43&              Y&       Y &        72   &           4540416&         2004-05-11&      48\ \ \ \ \      252\\
2487&    2885&    beta2 Tuc&       A2&      50&              Y&       Y &        72&              4540672&         2004-05-05&      48\ \ \ \ \      252\\
2578&    3003&    beta3 Tuc&       A0&      46&              Y&       Y &        72   &           4540928&         2004-06-22&      48\ \ \ \ \      504\\
2729&    3221&   &                 K5&      46&              Y&       Y &        72&              4541184&         2004-05-11&      192\ \ \ \      630\\
3556&   &    &                     M3&      38&              Y&  &        102     &        9022720&         2004-11-03&      48\ \ \ \ \      252\\
& &                CPD-64 120&      K1&      (73)&            Y& &      102&             9022464&         2004-05-11&      48\ \ \ \ \      N/A\\
6485&    8558&    &                G6&      49&              Y&       Y&         148    &         4813568&         2005-06-29   &   48\ \ \ \ \   126\\
6856&    9054&    CC Phe&          K1&      37&              Y&       Y &      149     &        4813824&         2004-11-03  &    48 \ \ \ \ \     126\\
9141&    12039&   DK Cet&          G4&      42&              Y&       Y&        148&             5310976&         2004-07-11 &     96\ \ \ \ \       252\\
9685&    12894&    &               F3&      47&              Y&  &        102&             9022208 &        2004-11-07&      48\ \ \ \ \      126\\
9892&    13183&    &               G6&      50&              Y&       Y&        152  &           4814592&         2004-11-07 &     48\ \ \ \ \      126\\
9902&    13246&    &               F7&      45&              Y&       Y&        153    &         4814848&         2004-11-07  &    48 \ \ \ \ \      126\\
&  &                CD-60 416&       K4&      (48)&            Y&       Y &       153&             4814848&         2004-11-07&      48\ \ \ \ \      126\\
10602&   14228&   phi Eri&         B8&      47&              Y&       Y&        102&             9021952&         2004-11-08   &   48\ \ \ \ \      252\\
12394&   16978&   eps Hyi&         B9&      47&              Y&       Y&       102&             9021184  &       2005-06-19&      48\ \ \ \ \      252\\
12413& 16754& & A1& 40& &  &      10&              3694848&         2005-07-30&      48\ \ \ \ \      126\\
&  &                CD-53 544&       K6&      (30)&            Y&       Y &       154&             4815104&         2004-11-07&      48\ \ \ \ \      126\\
&  &                AF Hor&          M2&      (27)&            Y&       Y&   154&             4815104&         2004-11-07&      48\ \ \ \ \      12\\
&  &                CD-58 553&       K5&      (50)&            Y?&      Y&   155&             4815360&         2004-11-07&      48\ \ \ \ \      126\\
14551& 19545& & A5& 58&  & &      10&              3696896&         2006-01-13&      48\ \ \ \ \      126\\
15247&   20385&      &             F5&      50&              Y&   &       3600   &         11257088&        2005-01-31   &   48\ \ \ \ \      252\\            
&  &               CD-46 1064&      K3&      50&    &       Y&      3600&            11257344&        2004-10-14&      48\ \ \ \ \      252\\
16449&   21997&   HR 1082&         A3&      73&    &                  Y*&        10&              3698432&         2004-09-24&      48\ \ \ \ \      126 \\
16853&   22705&   &                G2&      42&              Y&       Y &       3600&            11258112&        2004-10-14&      48\ \ \ \ \      252\\
17764&  24636& & F3& 55&  &  &   40566&           23051520&        2008-05-16&      96\ \ \ \ \      504\\
&  &                BD-15 705&       K3&      (63)&            Y&       Y*& 3600    &        11258880&        2005-02-01&      48\ \ \ \ \      278\\
21632&   29615&    &               G3&      55&              Y&       Y&       3600   &         11259392&        2005-03-04&      48\ \ \ \ \       252\\
21965&   30051&   &                F2&      58&              Y&       Y &       3600&            11259648&        2005-01-28&      48\ \ \ \ \      252\\
22295&   32195&   &                F7&      60&              Y& &       20707&           15009536&        2005-08-31&      96\ \ \ \ \      252 \\       
24947&   35114&   AS Col&          F6&      46&              Y&       Y*&       3600&            11260160&        2004-12-05&      48\ \ \ \ \      252\\
23179& 31647&  & A1& 49&  &  &  10&  3699712&  2004-02-25& 48\ \ \ \ \  126 \\
&  36869& &  G3& (59)&  & &      3600&            11260672&        2005-03-04&      48\ \ \ \ \      126\\
26309&  37286&  & A2& 57& &  &   10&              3700992&         2004-02-21&      48\ \ \ \ \      126\\
& &  AT Col&          K1&      (57)&            Y&       Y*&          3600&            11260928&        2005-02-28&      48\ \ \ \ \      278\\
26453&  37484& & F3& 60& &  &     148&             5307136&         2004-11-05&      96\ \ \ \ \      252 \\
&  &                CD-34 2406&      K4&      (60)&            Y&  &  3600&            11261440&        2005-02-28&      48\ \ \ \ \      278\\
26966&   38206&   HR 1975&         A0&      69&    &                  Y* &        40&              3983872&         2004-02-21&      48\ \ \ \ \      252\\
28036&   40216&     &              F7&      54&              Y&       Y* &   3600&            11261952&        2005-03-07&      48\ \ \ \ \      252\\
30030&   43989&   V1358 Ori&       G0&      50&              Y& &        148&             6598912&         2004-10-13&      96\ \ \ \ \      252 \\
30034&   44627&   AB Pic&          K2&      45&              Y&  &    3600 &           11262208   &     2004-12-25 &     48\ \ \ \ \      252\\
32104& 48097& &  A2& 43&  &  &        10&           3702784&         2004-03-15&      48\ \ \ \ \      126 \\
32235&   49855&   &                G6&      56&              Y&       Y*&       3600    &        11263488&        2004-11-08 &     48\ \ \ \ \      252\\
32435&   53842&   &                F5&      57&              Y& &    20707&           15003648 &       2005-08-31&      96\ \ \ \ \      76\\
33737&   55279&   &                K2&      64&              Y&       Y*&       3600&            11263744&        2005-03-08&      48\ \ \ \ \      278\\
83494& 154431& & A5& 54&  & &    10&              3721472&         2004-02-22&      48\ \ \ \ \       126\\
84642&  155915&  & G8& 55&  &  &     3600&            11266048&        2005-04-06&      48\ \ \ \ \       278\\
100751&  193924&  Peacock&         B2&      56&              Y& &      72&              4557312&         2004-04-09      & 48\ \ \ \ \      126 \\
104308&  200798&    &              A5&      66&              Y&  &      72      &        4558592 &        2004-09-23   &   96\ \ \ \ \      630 \\
105388&  202917&   &               G7&      46&              Y&       Y&      72   &           4558848  &       2004-09-24&      192\ \ \ \     630\\
105404&  202947&  BS Ind&          K0&      46&              Y& &      72&              4559104     &    2004-09-24   &   192\ \ \ \      630 \\
107345&  &      &                  M0&      42&              Y&       Y&      72      &        4559360&         2004-09-23 &     498\ \ \ \     630 \\
107947&  207575&  &                  F6&      45&              Y&       Y&      72   &           4560128   &      2004-09-23 &     498\ \ \ \      630\\
108195&  207964&  HR 8352&         F1&      46&              Y&       Y&   72&              4560384     &    2004-09-23 &     498\ \ \ \      630\\
108422&  208233&    &              G9&      55&              Y&  &  3600&            11266304 &       2004-11-06&      48\ \ \ \ \       252\\
116748&  222259&  DS Tuc&          G6&      46&              Y&  &      72    &          4561920 &        2204-05-05&      192\ \ \ \     630 \\
118121&  224392&  eta Tuc&         A1&      49&              Y&  &      72    &          4562176 &        2004-06-22   &   48\ \ \ \ \      630 \\
\hline
\end{longtable}
\noindent Note $-$  Distances in parenthesis are estimated photometrically.  Y* in the T08 column
indicates that Torres et al (2008) place the star in what they designate as
either the Columba or Carina Association.  Because these
associations have the same age and nearly the same UVW space motions as
Tucana/Horologium, for the purposes of Table 11 of the present paper we consider as a
single age group the stars in all of these associations.  Spitzer observations of HIP 114189 (= HR 8799, Table 3) have been analyzed in detail by Su et al (2009) and we have therefore not included this star in Table 6. 
\end{landscape}

\clearpage

\begin{table}
\caption{Spitzer Observational Programs That Include Nearby Argus Stars}
\begin{tabular}{@{}lcccc}
\br
Star& Spitzer& AORKEY& Date of& Integ. time (sec)\\
& Program ID& Archive ID& Observation& 24$\mu$m \ 70$\mu$m\\
\mr
HD 84075& 72& 4545280& 2004-02-23& 288\ \ \ \ \  630\\
HIP 50191&  10&  3707136&  2005-05-18&  48\ \ \ \ \ 126\\
CD-74 673& 148& 5355520&  2004-08-23& 288\ \ \ \ \ 756\\
HIP 68994&  20597&  15591424&  2006-04-08&  30\ \ \ \ \ 15\\
HIP 79797& 10& 3720448& 2004-03-17& 48\ \ \ \ \ 126\\
HIP 98495& 10&  3725824&  2004-04-08&  48\ \ \ \ \  126\\
HIP 99770&  10&  3726848&  2004-10-16&  48\ \ \ \ \ 126 \\
\br
\end{tabular}
\end{table}
\noindent Note $-$ The infrared spectrum of HIP 57632 (= $\beta$ Leo, Table 4) has been extensively studied (beginning with IRAS, Rhee et al 2007a) and we therefore have not included this star in Table 7.

\clearpage

\def\mc{\multicolumn}
\def\noIR{\tiny no IR source}
\def\str{\tiny Strange spec.}
\begin{landscape}
\begin{longtable}{lcclcccccc}
\caption{AB Dor Stars: MIPS flux densities}\\
\hline
Star& photosph.\ \  meas.& Excess?& photosph.\ \ meas.& Excess?& IRS?& T$_{star}$& R$_{star}$& T$_{dust}$& L$_{dust}$/L$_{*}$ \\
 & \ \ \ \ \ \  24 $\mu$m (mJy)&  & \ \ \ \ 70$\mu$m (mJy)&  &  & (K)& (R$_{\odot}$)& (K)& (x10$^{-5}$)\\
\hline
\hline
\endfirsthead
\hline
\caption[]{(continued)}\\
Star& photosph.\ \  meas.& Excess?& photosph.\ \ meas.& Excess?& IRS?& T$_{star}$& R$_{star}$& T$_{dust}$& L$_{dust}$/L$_{*}$ \\
 & \ \ \ \ \ \  24 $\mu$m (mJy)&  & \ \ \ \ 70$\mu$m (mJy)&  &  & (K)& (R$_{\odot}$)& (K)& (x10$^{-5}$)\\
\hline
\hline
\hline
\endhead
\hline
\hline
\multicolumn{5}{r}{\small\sl continued on next page}\\
 \hline
 \endfoot
 \hline
 \endlastfoot
PW And& 20.6\ \ \ \ \ \ 21.8& N& \ \ \   2.3& N&  & 4700& 0.54& & \\
HIP 3589& 19.9\ \ \ \ \ \ 21.9& N& \ \ \   2.2& N&  & 6000& 1.19& & \\
HIP 5191& 8.4\ \ \ \ \ \  8.62& N& \ \ \    0.9& N& & 4900& 0.87& & \\
HIP 6276& 16.3\ \ \ \ \ \  19.8&   Y&\ \ \       1.8&                N&       Y& 5200& 0.84& 135& 5.2 \\
HIP 10272&        35.7\ \ \ \ \ \   37.0&   N&\ \ \       3.9&                N& & 5200& 1.14& & \\      
HIP 13027A&       46.0\ \ \ \ \ \   26.7&    N&\ \ \ 5.0&                N& & 5200& 1.31& & \\      
HIP 13027B&       39.2\ \ \ \ \ \    26.7&   N&  \ \ \     4.3&                N& & 5000& 1.23&  & \\
IS Eri&          13.7\ \ \ \ \ \   18.8&    Y&\ \ \       1.5&                N&       Y& 5200& 0.88& 150&  9.9  \\ 
HIP 14807&        6.2\ \ \ \ \ \   7.57&    Y&\ \ \       0.7&                N& & 4600& 0.79&  $>$70& $<$37 \\      
HIP 14809&        11.5\ \ \ \ \ \ 16.1&    Y&\ \ \       1.3&                N&  & 5900& 0.91&  $>$97& $<$10 \\ 
HIP 15353&        37.1\ \ \ \ \ \   35.8&    N&\ \ \       4.3&                N& & 8100& 1.62& &  \\       
HIP 16563A&       19.1\ \ \ \ \ \   25.0&    Y?*&\ \ \       2.1&          N& & 5400& 0.86& $>$115& $<$7.9 \\      
HIP 16563B&       6.6\ \ \ \ \ \ 9.31&    Y?*&\ \ \       0.7&              N& & 4400& 0.54& $>$93& $<$32 \\      
HIP 17695&        19.6\ \ \ \ \ \  17.6& N&\ \ \       2.3&                 N& & 3400& 0.49&  & \\
HIP 18859&        154\ \ \ \ \ \   210&    Y&\ \ \       18.0 \ \ \ 323&     Y&       Y& 6200& 1.24& 70& 11 \\
HIP 19183&        16.1\ \ \ \ \ \    14.5&    N&\ \ \       1.9&                N&       Y& 6400& 1.13&  & \\
HIP 25283&        37.5\ \ \ \ \ \   35.3&    N&\ \ \       4.3 \ \ \ \ 38.3&     Y&  & 4000& 0.70& 60& 17.5  \\    
AB DorA&         101\ \ \ \ \ \   111&    N&\ \ \       11.1&                N& & 4800& 0.90& & \\      
AB DorB&         7.2\ \ \ \ \ \   7.01&    N&\ \ \       0.8&                N& & & & & \\      
HIP 26369&        16.0\ \ \ \ \ \   19.5&    Y&\ \ \       1.8&                N& & 4500& 0.60& $>$120& $<$9.5 \\      
HIP 26373&        31.5\ \ \ \ \ \   33.3&    N&\ \ \       3.4&                N&       Y& 5200& 0.79& & \\
CD-35 2722&      16.6\ \ \ \ \ \    14.4&    N&\ \ \       2.0&                N& & 3600&  0.65& &\\ 
HIP 30314&        63.3\ \ \ \ \ \   70.9&    N&\ \ \       6.9&                N&       Y& 6000& 1.02&  &  \\         
GSC8894-0426&   17.6\ \ \ \ \ \   13.6&    N&\ \ \       2.1&                N&  & 3200& 0.58& & \\   
HIP 31711&        108\ \ \ \ \ \   117&    N&\ \ \       11.6&                N&   & 5700& 1.27& &    \\
HIP 31878&        19.2\ \ \ \ \ \   20.2&    N&\ \ \       2.2&                N& &  4300& 0.60& & \\      
HIP 36349&        65.0\ \ \ \ \ \   49.8&    N&\ \ \       7.7&                N&   & 3400& 0.85& &     \\
HIP 51317&      86.3\ \ \ \ \ \   73.7&    N&\ \ \       10.1&                N&  & 3500& 0.45& & \\ 
BD+20 1790&      13.1\ \ \ \ \ \    12.2&    N&\ \ \       1.5&                N&       Y& 4400& 0.58&  &  \\
HIP 63742&        41.6\ \ \ \ \ \   48.4&    Y&\ \ \       4.4&             N&  & 5200&  0.83& $>$120& $<$4.5 \\
HIP 76768&        13.7\ \ \ \ \ \    13.6&    N&\ \ \       1.6&                N& & 4100& 1.00&  & \\      
HIP 81084&        9.9\ \ \ \ \ \   8.15&    N&\ \ \       1.2&                N&  & 3700& 0.66&  & \\    
HIP 82688&        19.2\ \ \ \ \ \    20.9&    N&\ \ \       2.1&                N&       Y& 6000& 1.14& &  \\
HD 317617&        6.2\ \ \ \ \ \   5.49&    N&\ \ \       0.7&                N&  & 4700& 0.83&  & \\
HIP 86346&        15.0\ \ \ \ \ \   16.7&    N&\ \ \       1.7&                N&  & 4200& 0.60&  &      \\
HIP 93580&        52.6\ \ \ \ \ \    55.7&    N&\ \ \       6.1&                N& &  7700&  1.88&  & \\ 
HIP 106231&       21.5\ \ \ \ \ \   23.5&    N&\ \ \       2.4&                N&  &  4500& 0.71& &       \\
HIP 107948&       14.7\ \ \ \ \ \   12.6&    N&\ \ \       1.7&                N&  & 3300&  0.84&  &  \\      
HIP 109268&       951\ \ \ \ \ \   970&    N&\ \ \       104&                N& & 12500& 3.77&  & \\   
HIP 110526&       48.8\ \ \ \ \ \   40.9&    N&\ \ \       5.8&                N& & 3200& 0.78&  &  \\      
HIP 113579&       29.0\ \ \ \ \ \   31.4&    N& \ \ \      3.2&                N&       Y& 5900& 0.94& &  \\
HIP 113597&       31.2\ \ \ \ \ \   26.2&    N&\ \ \       3.6&                N& & 3800& 1.09&  & \\      
HIP 114066&       12.3\ \ \ \ \ \    11.1&    N&\ \ \       1.4&                N&  & 4100&  0.56&  &  \\      
HIP 114530&       10.3\ \ \ \ \ \   12.3&    Y&\ \ \       1.1&            N& &  5600& 0.92&  $>$85& $<$7.2 \\     
HIP 115738*&       78.2\ \ \ \ \ \   111&    Y&\ \ \       8.9&              N&  Y& 9000& 1.93&  &  \\
HIP 117452&       104\ \ \ \ \ \   168&    Y&\ \ \       11.8\ \ \ \  49.7&     Y&  Y&  9200& 1.96& 170& 2.9  \\
HIP 118008&       28.9\ \ \ \ \ \   38.6&    Y&\ \ \      3.2&          N&  &  5000& 0.71& $>$140& $<$8.9 \\    
\hline
\end{longtable}
\noindent Notes $-$ *See discussion in Section 5.5.1.  The second and fifth columns give the expected photospheric fluxes (see
Section 3) and the third and sixth columns give the MIPS measured fluxes.
For stars in Figures 1 and 2 where MIPS measured only an upper limit to the 70 $\mu$m flux density, the above tabulated L$_{dust}$/L$_{*}$ is an upper limit and T$_{dust}$ is a lower limit.  
\end{landscape}

\clearpage

\def\mc{\multicolumn}
\def\noIR{\tiny no IR source}
\def\str{\tiny Strange spec.}
\begin{landscape}
\begin{longtable}{lcclcccccc}
\caption{Tuc/Hor \& Columba Stars: MIPS flux densities} \\
\hline
Star& photosph.\ \  meas.& Excess?& photosph.\ \ meas.& Excess?& IRS?& T$_{star}$& R$_{star}$& T$_{dust}$& L$_{dust}$/L$_{*}$ \\
 & \ \ \ \ \ \  24 $\mu$m (mJy)&  & \ \ \ \ 70$\mu$m (mJy)&  &  & (K)& (R$_{\odot}$)& (K)& (x10$^{-5}$)\\
\hline
\hline
\endfirsthead
\hline
\caption[]{(continued)}\\
Star& photosph.\ \  meas.& Excess?& photosph.\ \ meas.& Excess?& IRS?& T$_{star}$& R$_{star}$& T$_{dust}$& L$_{dust}$/L$_{*}$ \\
 & \ \ \ \ \ \  24 $\mu$m (mJy)&  & \ \ \ \ 70$\mu$m (mJy)&  &  & (K)& (R$_{\odot}$)& (K)& (x10$^{-5}$)\\
\hline
\hline
\hline
\endhead
\hline
\hline
\multicolumn{5}{r}{\small\sl continued on next page}\\
 \hline
 \endfoot
 \hline
 \endlastfoot
HIP 490& 25.3\ \ \ \ \ \ 27.5& N& \ \ \ 3.0\ \ \ \ \ \ 162& Y& Y& 6100& 1.06& 50& 34.0  \\
HIP 1113& 10.8\ \ \ \ \ \ 12.7& Y& \ \ \ 1.1\ \ \ \ \ \ 18.7& Y& & 5600& 0.82&  70& 11.9   \\
HIP 1481& 24.9\ \ \ \ \ \ 35.2& Y& \ \ \ 2.9\ \ \ \ \ \ 14.3& Y& Y&  6200& 1.06& 132&  7.1 \\
HIP 1910& 9.8\ \ \ \ \ \ 9.03& N&\ \ \ 1.1& N& & 3800& 0.94& & \\
HIP 1993& 8.2\ \ \ \ \ \ 7.68& N& \ \ \ 0.96& N&  & 3700& 0.70&  & \\ 
HIP 2484& 97.3\ \ \ \ \ \ 87.8& N& \ \ \ 10.4& N& Y& 11500& 1.72&  & \\
HIP 2487& 158\ \ \ \ \ \ \ 140& N& \ \ \ 18.3& N& &  8200&  & & \\
HIP 2578& 68.3\ \ \ \ \ \  231& Y& \ \ \ 7.7\ \ \ \ \ \ 59.9& Y& Y& 9300& 1.67& 200& 9.4  \\
HIP 2729& 18.1\ \ \ \ \ \  17.1& N& \ \ \ 2.1& N& & 4400& 1.21&  & \\
HIP 3556& 9.6\ \ \ \ \ \ 7.71& N& \ \ \ 1.1& N& & 3500& 0.80&  & \\
CPD-64 120& 4.1\ \ \ \ \ \ 5.28& Y&\ \ \ 0.44& &  &  5200& 0.88&  & \\
HD 8558& 12.0\ \ \ \ \ \ 14.3& Y& \ \ \ 1.3& N&  &  5800& 0.95&  $>$75&  $<$ 9.0  \\
HD 9054& 12.7\ \ \ \ \ \ 14.1& N& \ \ \  1.4& N&  & 4900& 0.80&  & \\
HIP 9141& 17.3\ \ \ \ \ \ 24.7& Y& \ \ \ 1.9& N& Y&  5800& 0.98& 160& 7.9 \\
HD 12894& 47.0\ \ \ \ \ \ 48.6& N& \ \ \ 5.5& N&  & 6800& 1.61&  & \\
HD 13183& 12.5\ \ \ \ \ \  14.1& Y?& \ \ \ 1.35& N&  & 5400& 1.03&  & \\
HD 13246& 23.7\ \ \ \ \ \ 47.3& Y& \ \ \ 2.76 \ \ \ \ 31.7& Y& Y&  6200& 1.13&  125& 17.0  \\
CD-60 416& 6.8\ \ \ \ \ \  8.30& N&\ \ \  0.76&  N&  Y&  4600& 0.77&  & \\
HD 14228& 176\ \ \ \ \ \  188& N& \ \ \  19.1& N& & 12000& 2.53&  &  \\
HIP 12394& 129\ \ \ \ \ \ 132& N& \ \ \ 14.0& N& & 10500& 2.23& & \\
HIP 12413& 114\ \ \ \ \ \ \ 117& N& \ \ \ 13.2& N& & 8300&  1.93&  & \\
CD-53 544& 15.1\ \ \ \ \ \ 15.0& N&\ \ \  1.7&   N&  & 4200&  0.74&  & \\
AF Hor& 9.77\ \ \ \ \ \ 7.94& N&\ \ \ 1.14& N&   &  3500& 0.57&  & \\
CD-58 553& 5.7\ \ \ \ \ \  5.35& N&\ \ \ 0.65& N&  & 4400& 0.74&  & \\
HIP 14451& 34.0\ \ \ \ \ \ 34.2& N& \ \ \ 3.9& N& &  8000& 1.54&  & \\
HIP 15247& 25.4\ \ \ \ \ \  27.5& N& \ \ \ 3.0& N&  & 6200& 1.10&  & \\
CD-46 1064& 10.2\ \ \ \ \ \ 11.8& Y?& \ \ \ 1.1& N& & 4800&  0.95&  & \\
HD 21997& 27.0\ \ \ \ \ \ 54.6& Y& \ \ \  2.9\ \ \ \ \ \ 70.5& Y& Y& 8200& 1.74&  60&  57.6  \\
HIP 16853& 23.3\ \ \ \ \ \  27.2& Y& \ \ \ 2.5& N& & 6000& 1.10& $>$95& $<$2.8  \\
HIP 17764& 24.8\ \ \ \ \ \ 42.6& Y& \ \ \ 2.9\ \ \ \ \ \ 36.0& Y& Y& 6700& 1.37& 110& 11.4  \\
BD-15 705& 7.3\ \ \ \ \ \ 7.43& N& \ \ \ 0.83& N& & 4600& 1.05&  &  \\
HIP 21632& 12.2\ \ \ \ \ \ 12.5& N& \ \ \ 1.3& N& & 5900& 1.03& & \\
HIP 21965& 28.1\ \ \ \ \ \  29.7& N& \ \ \ 3.3\ \ \ \ \ \ 25.4& Y& & 6600&  1.55&  80& 4.4 \\
HIP 22295& 12.6\ \ \ \ \ \ 16.4& Y& \ \ \ 1.5\ \ \ \ \ \ 11.3& Y& &  6300&  1.10&  100& 7.2 \\
HIP 23179*& ??\ \ \ \ \ \ \  92.6& N& \ \ \ $\sim$10& N& Y& 8000& 2.30&  & \\
HIP 24947& 25.4\ \ \ \ \ \ 32.6& Y& \ \ \ 2.9\ \ \ \ \ \ 11.0& Y&  & 6300& 1.18& 135& 4.9  \\
HD 36869& 13.6\ \ \ \ \ \ 14.0& N& \ \ \ 1.5& N& & 5800& 1.21&  &\\
HIP 26309& 31.3\ \ \ \ \ \  67.9& Y& \ \ \ 3.6& N& Y& 8000& 1.46& 160& 8.2   \\
AT Col& 7.7\ \ \ \ \ \ 9.54& Y& \ \ \ 0.84& N& & 5100& 0.95&  &  \\
HIP 26453& 22.3\ \ \ \ \ \ 56.1& Y& \ \ \ 2.6\ \ \ \ \ \ 116& Y& Y& 6700& 1.40& 90& 32.5  \\
CD-34 2406& 2.05\ \ \ \ \ \  1.70& N& \ \ \ 0.23& N& & 4800& 0.52&  &\\
HD 38206& 32.5\ \ \ \ \ \  114& Y& \ \ \ 3.6\ \ \ \ \ \ 388& Y& Y& 9900& 1.68&  85& 19.1  \\
HIP 28036& 23.8\ \ \ \ \ \ 23.7& N& \ \ \ 2.8& N& & 6300& 1.36& & \\
HIP 30030& 17.3\ \ \ \ \ \ 20.4& Y& \ \ \ 2.0& N& Y& 6100& 1.08&  160& 3.2 \\
HIP 30034$^a$& 10.4\ \ \ \ \ \ 13.1& Y& \ \ \ 1.1& N& & 5200& 0.87&  & \\
HIP 32104& 67.4\ \ \ \ \ \  66.7& N& \ \ \ 7.8& N& & 8600& 1.57&  &\\
HIP 32235& 7.9\ \ \ \ \ \  9.81& Y& \ \ \ 0.86& N& & 5800& 0.89&  $>$75&  $<$13.2  \\
HIP 32435$^a$& 21.5\ \ \ \ \ \  33.2& Y& \ \ \ 2.5& & Y& 6500& 1.35& 140& 7.4   \\
HIP 33737& 5.9\ \ \ \ \ \  6.29& N& \ \ \ 0.66& N& &  4900& 0.94&  & \\
HIP 83494& 41.5 \ \ \ \ \ \ 47.3& Y?& \ \ \ 4.8& N& & 7700& 1.63& & \\
HIP 84642& 6.75\ \ \ \ \ \ 7.87& Y& \ \ \ 0.73& N& & 5200& 0.84&  & \\
HIP 100751& 680\ \ \ \ \ \ 656& N& \ \ \ 78.4& N& &  15000& 5.25&  & \\
HIP 104308& 26.4\ \ \ \ \ \ 25.7& N& \ \ \ 3.1& N& & 7600& 1.61& & \\
HIP 105388& 11.1\ \ \ \ \ \  20.1& Y& \ \ \ 1.2 \ \ \ \ \ 36.3& Y& Y& 5700& 0.86&  86& 27.7  \\
HIP 105404& 15.2\ \ \ \ \ \ 19.5& Y& \ \ \ 1.7& N& &  5100& 1.07&  $>$120&  $<$8.6   \\
HIP 107345& 6.5\ \ \ \ \ \ 6.44& N& \ \ \ 0.76& N& & 3900& 0.69&  & \\
HIP 107947& 27.2\ \ \ \ \ \ 31.0& Y?& \ \ \ 3.2\ \ \ \ \ \  9.6& Y& & 6400& 1.20& 110& 2.4 \\
HIP 108195& 80.5\ \ \ \ \ \ 77.3& N& \ \ \ 9.4& N& & 6600& 2.1&  & \\
HIP 108422& 13.2\ \ \ \ \ \  17.3& Y& \ \ \ 1.4& N& &  5200& 1.18&  $>$100& $<$12  \\
HIP 116748*& 25.0\ \ \ \ \ \  24.8& N& \ \ \ 2.7& N& &  5200& 1.37&  & \\
HIP 118121& 78.7\ \ \ \ \ \ 77.8& N& \ \ \ 9.0& N& & 9000& 1.90&  & \\
\hline
\end{longtable}
\noindent Notes $-$  $^a$See discussion of HIP 30034 and HIP 32435 in Section 5.5.1.  *HIP 23179 and 116748 are each 5" binary stars.  Flux densities for HIP 116748 pertain to the sum of the primary and secondary.  Due to relatively large errors for the 2MASS near-IR magnitudes for the HIP 23179 secondary, an accurate estimate of the total 24 $\mu$m photospheric flux density of this binary system is not now possible.  However, there is no indication of excess mid-IR emission in the IRS spectrum out to $\sim$35 $\mu$m wavelength.   Spitzer observations of HIP 114189 (= HR 8799, Table 3) have been analyzed in detail by Su et al (2009); therefore we have not included this star in Table 9.  HIP 114189 has excess emission at both 24 and 70 $\mu$m and it is included in the statistics in Table 11.  For stars in Figures 3-6 where MIPS measured only an upper limit to the 70 $\mu$m flux density, L$_{dust}$/L$_{*}$ tabulated in Table 9 is an upper limit and T$_{dust}$ is a lower limit.

\clearpage

\begin{table}
\caption{Argus Stars: MIPS flux densities}
\begin{tabular}{@{}lcclcccccc}
\br
Star& photosph.\ \  meas.& Excess?& photosph.\ \ meas.& Excess?& IRS?& T$_{star}$& R$_{star}$& T$_{dust}$& L$_{dust}$/L$_{*}$ \\
 & \ \ \ \ \ \  24 $\mu$m (mJy)&  & \ \ \ \ 70$\mu$m (mJy)&  &  & (K)& (R$_{\odot}$)& (K)& (x10$^{-5}$)\\
\mr
HD 84075& 9.0\ \ \ \ \ \ 12.6& Y& \ \ \ 1.0\ \ \ \ \ \ \ 35.8& Y& Y& 6000& 1.03& 170& 25.1 \\
HIP 50191& 215\ \ \ \ \ \ 266& Y& \ \ \ 24.4\ \ \ \ \ \  46.2& Y& &  9000& 2.03& 180& 1.0 \\
CD-74 673& 4.6\ \ \ \ \ \  4.85& N& \ \ \ 0.51& N& Y&  4800& 0.64& & \\
HIP 68994$^a$& 14.8\ \ \ \ \ \ 24.7& Y& \ \ \ 1.7& & &  6600& 1.28&  & \\
HIP 79797& 38.4\ \ \ \ \ \ 52.6& Y& \ \ \ 4.4\ \ \ \ \ \ \ 39.3& Y& Y& 8200& 1.56& 220& 5.1 \\
HIP 98495& 175\ \ \ \ \ \  173& N& \ \ \ 19.7\ \ \ \ \ \ 57.8& Y& Y& 9600& 1.85& 90&  0.44 \\
HIP 99770& 121\ \ \ \ \ \ 125& N& \ \ \ 13.9\ \ \ \ \ \ \ * & N?& & 7600& 2.12& & \\
\br
\end{tabular}
\end{table}
\noindent Note $-$ $^a$See discussion of HIP 68994 in Section 5.5.1.  *HIP 99770 is located only 1 degree from the Galactic plane and the MIPS 70 $\mu$m image is messy, containing widespread extended emission.  Therefore, although our formal aperture photometry indicates a 70 $\mu$m flux density of 192 mJy, we regard this "detection" as uncertain.  The infrared spectrum of HIP 57632 (= $\beta$ Leo, Table 4) has been extensively studied (beginning with IRAS, e.g., Rhee et al 2007a) and we therefore have not included this star in Table 10.  HIP 57632 has excess emission at both 24 and 70 $\mu$m and it is included in the statistics in Table 11.
\end{landscape}

\clearpage

\begin{table}
\caption{Infrared Excess Fractions in Clusters/Associations}
\begin{tabular}{@{}lcccc}
\br
Cluster/& age& 24 $\mu$m excess& 70 $\mu$m excess& Reference\\
Association& (Myr)& (\#) \ \ \ \ \ (\%)& (\#) \ \ \ \ \ (\%)& \\
\mr
$\eta$ Cha& 6*& 9/16   \ \ \ \ 56& \ 5/15  \ \ \ \ $>$33& Rebull 2008\\
TW Hya Assoc. & 8& 7/23 \ \ \ \ 30& \ 6/20 \ \ \ \ $>$30& Rebull 2008 \\
UCL/LCC& 10& 10/35 \ \ \ 34& \ 7/35 \ \ \ \ $>$20& Rebull 2008\\
$\beta$ Pic MG& 12& 7/30 \ \ \ \ 23& \ 11/30 \ \ \ \ $>$37& Rebull 2008\\
NGC 2547& 30& 16/38 \ \ \ 42& & Gaspar 2009\\
Tuc/Hor/Columba& 30& 27/62 \ \ \ 43.5& \ 15/60 \ \ \ \ $>$25& this paper\\
IC 2391& 40*& 6/26 \ \ \ \ 23&  & Rebull \& Gaspar\\
Argus Assoc. & 40*& 5/8 \ \ \ \ \ 62.5& \ 5/7 \ \ \ \ $>$71& this paper\\
NGC 2451& 60& 12/38 \ \ \ 32&  & Balog 2009\\
AB Dor MG& 70& 12/47 \ \ \ 25.5& \ 3/47 \ \ \ \ $>$6.4& this paper\\
Pleiades& 100& 10/73 \ \ \ \ 14& & Gorlova 2006\\
M47& 100& 8/63 \ \ \ \ 13& & Rebull 2008\\
Hyades& 650& 2/78 \ \ \ \ 2.5&  & Gaspar 2009\\
Praesepe& 750& 1/135 \ \ \ \ 0.7&  & Gaspar 2009\\ 
\br
\end{tabular}
\end{table}
\noindent Note $-$ * age from Torres et al. (2008)

\clearpage
\begin{figure}
\includegraphics[width=140mm]{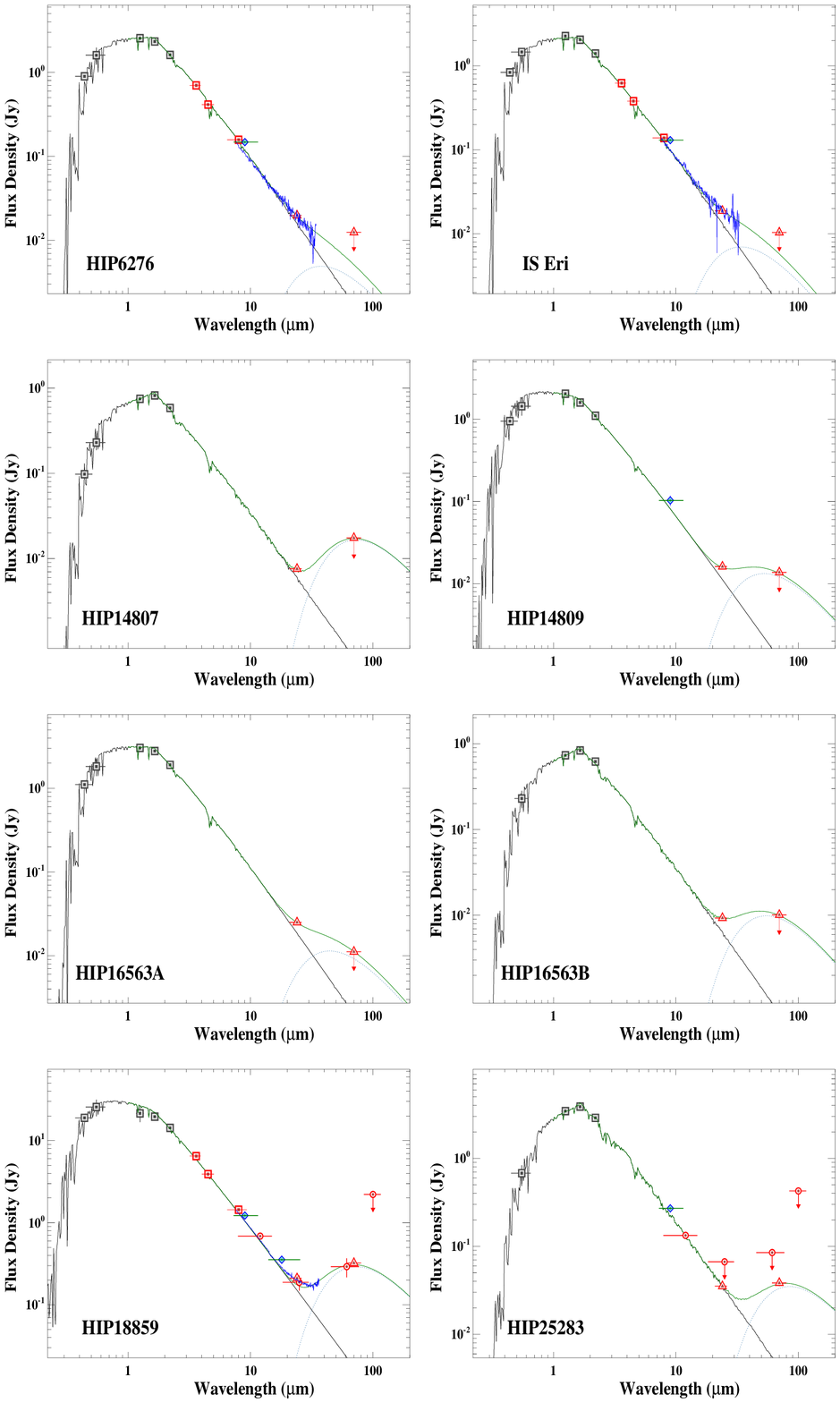}
\caption{\label{figure1}}
\end{figure}

\clearpage
Spectral energy distributions (SEDs) for AB Dor stars in Table 8 with probable or definite infrared excess emission.  Near infrared JHK data points are from the 2MASS catalog.  Square data points between 3.5 and 8 $\mu$m are from the IRAC camera on Spitzer.  Diamonds at 9 and 18 $\mu$m are from AKARI.  Triangles at 24 and 70 $\mu$m are from the MIPS camera on Spitzer.  Circles between 12 and 100 $\mu$m are from IRAS.  The IRS spectrum from Spitzer is plotted at mid-IR wavelengths.

\clearpage
\begin{figure}
\includegraphics[width=140mm]{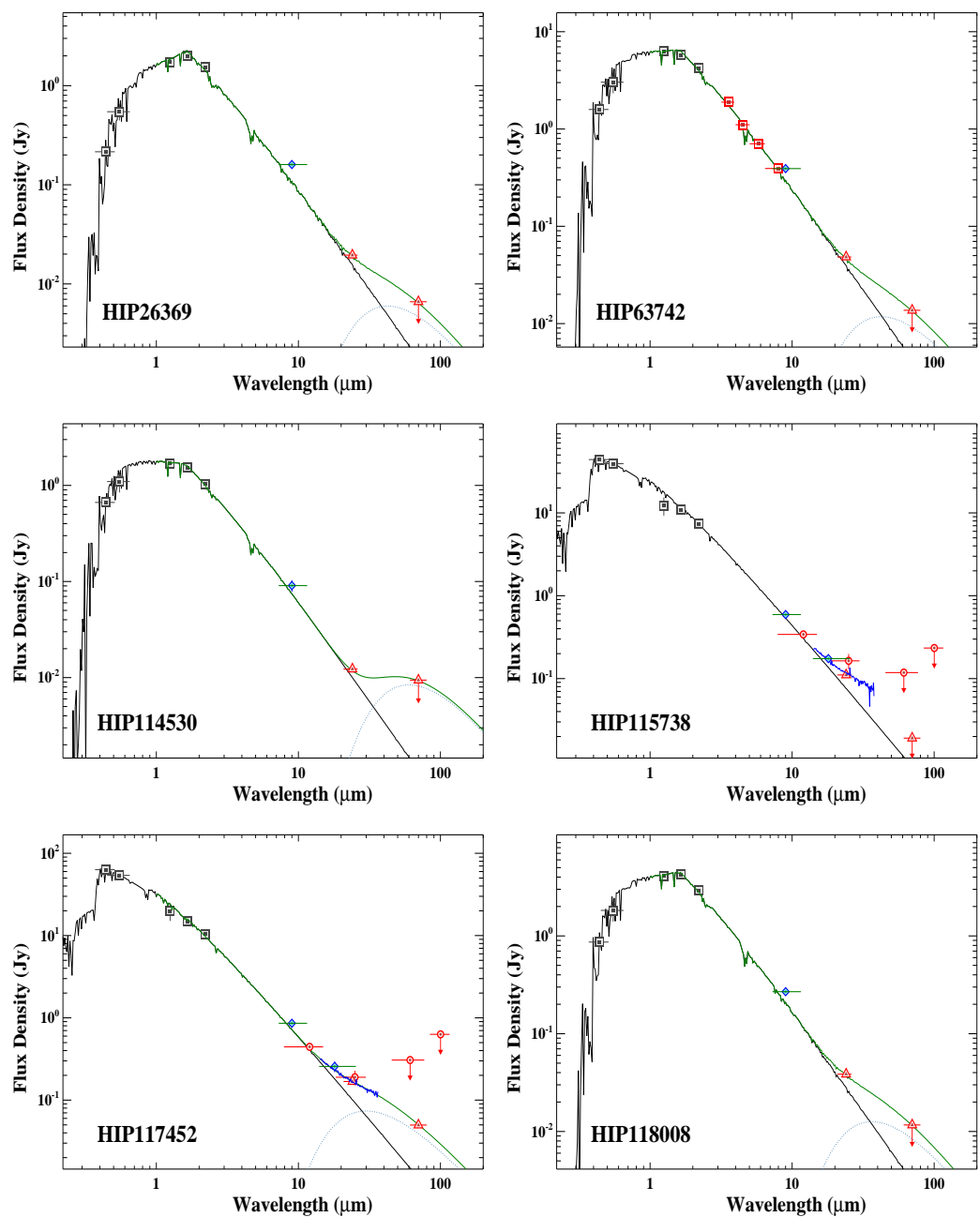}
\caption{\label{figure2} Spectral energy distributions (SEDs) for AB Dor stars in Table 8 with probable or definite infrared excess emission.}
\end{figure}

\clearpage
\begin{figure}
\includegraphics[width=140mm]{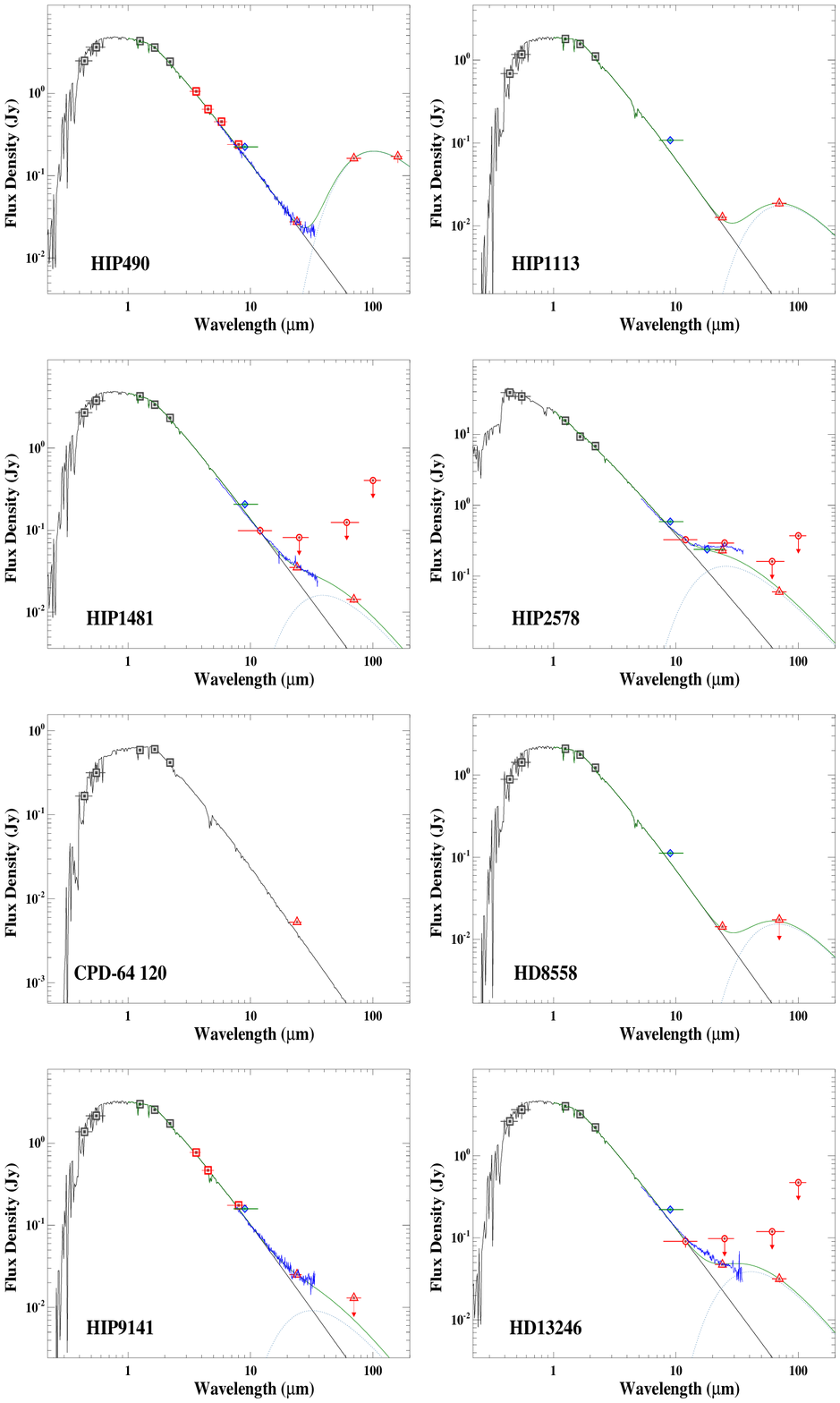}
\caption{\label{figure3} Spectral energy distributions (SEDs) for Tuc/Hor stars in Table 9 with probable or definite infrared excess emission.}
\end{figure}

\clearpage
\begin{figure}
\includegraphics[width=140mm]{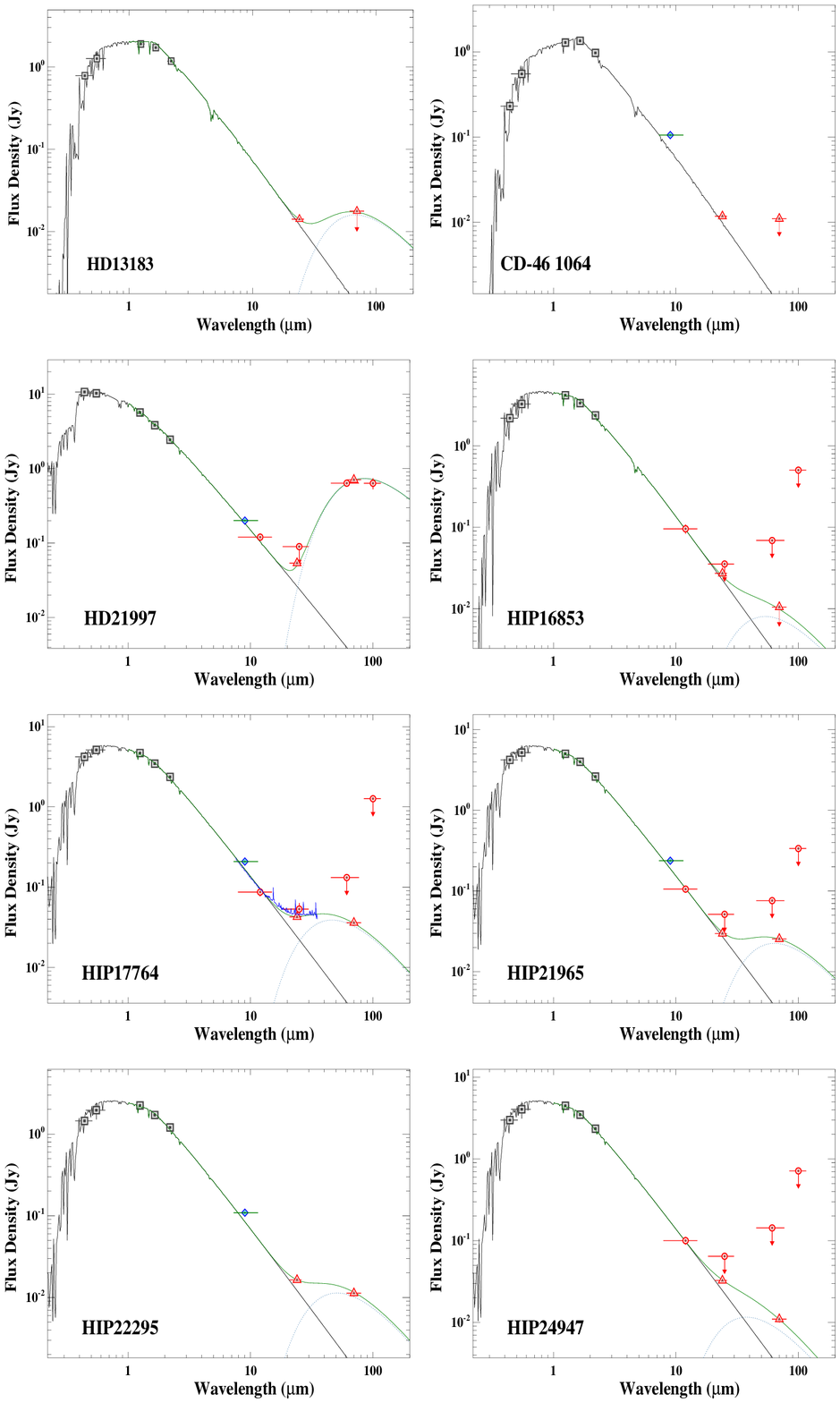}
\caption{\label{figure4} Spectral energy distributions (SEDs) for Tuc/Hor stars in Table 9 with probable or definite infrared excess emission.}
\end{figure}

\clearpage
\begin{figure}
\includegraphics[width=140mm]{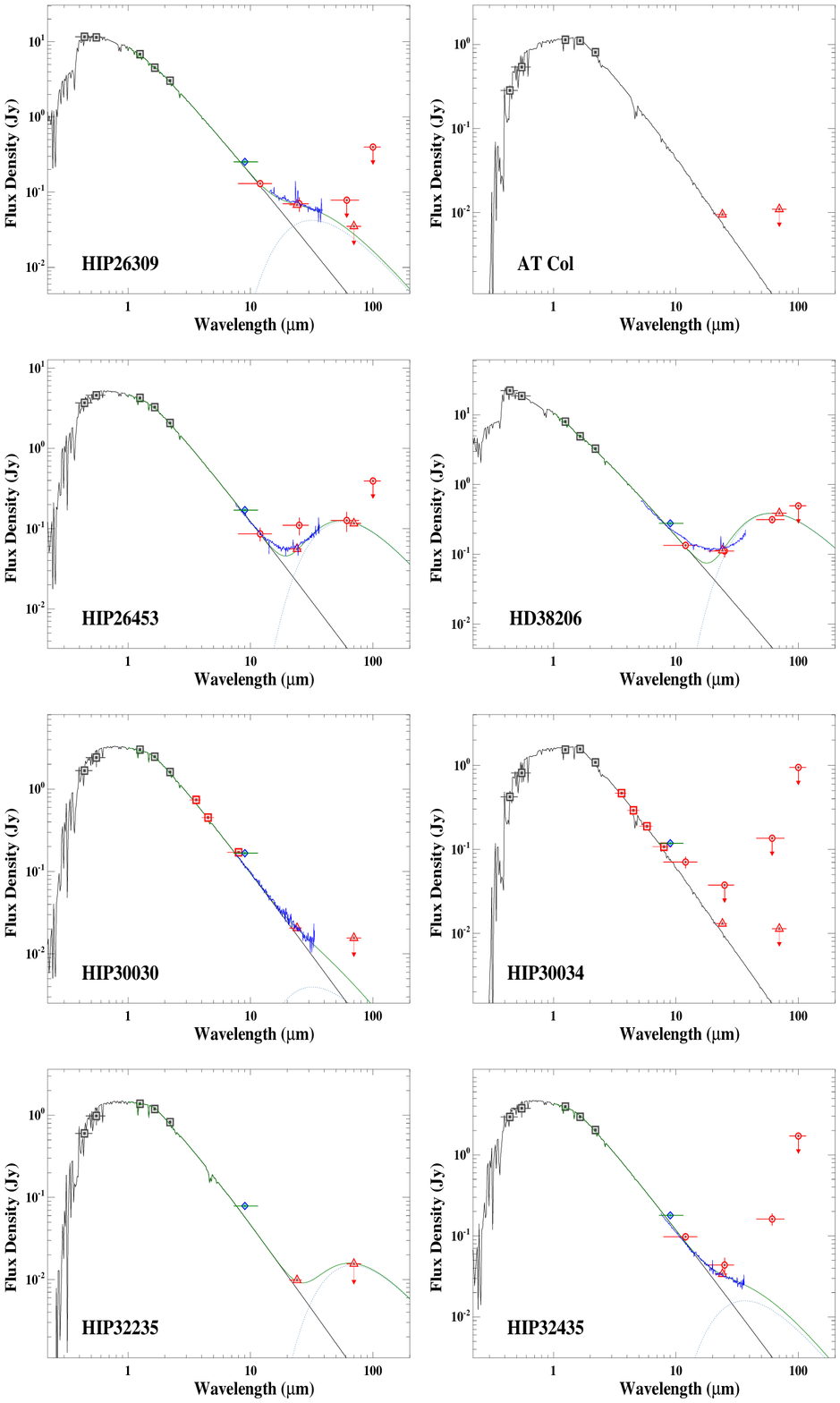}
\caption{\label{figure5} Spectral energy distributions (SEDs) for Tuc/Hor stars in Table 9 with probable or definite infrared excess emission.}
\end{figure}

\clearpage
\begin{figure}
\includegraphics[width=140mm]{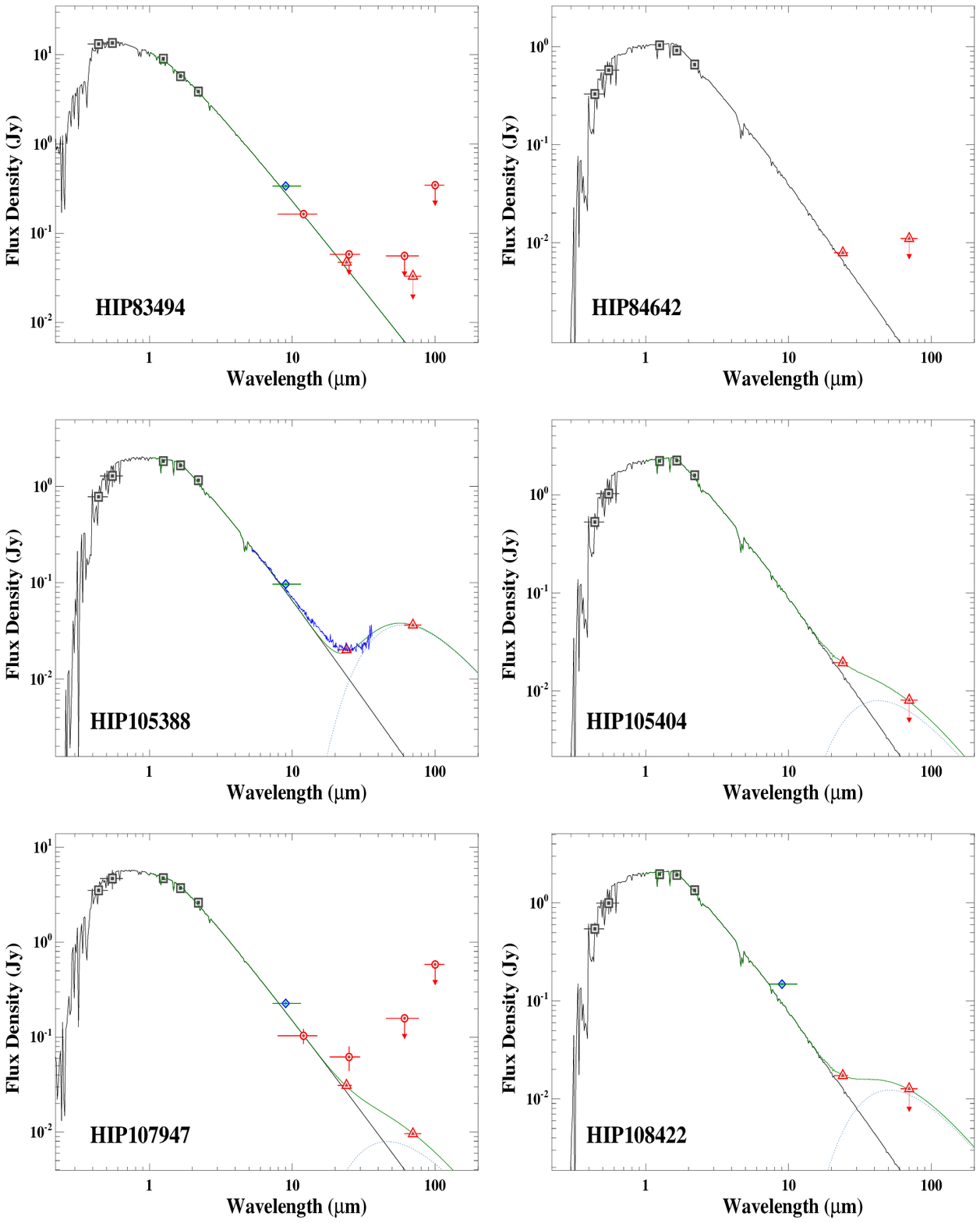}
\caption{\label{figure6} Spectral energy distributions (SEDs) for Tuc/Hor stars in Table 9 with probable or definite infrared excess emission.}
\end{figure}

\clearpage
\begin{figure}
\includegraphics[width=140mm]{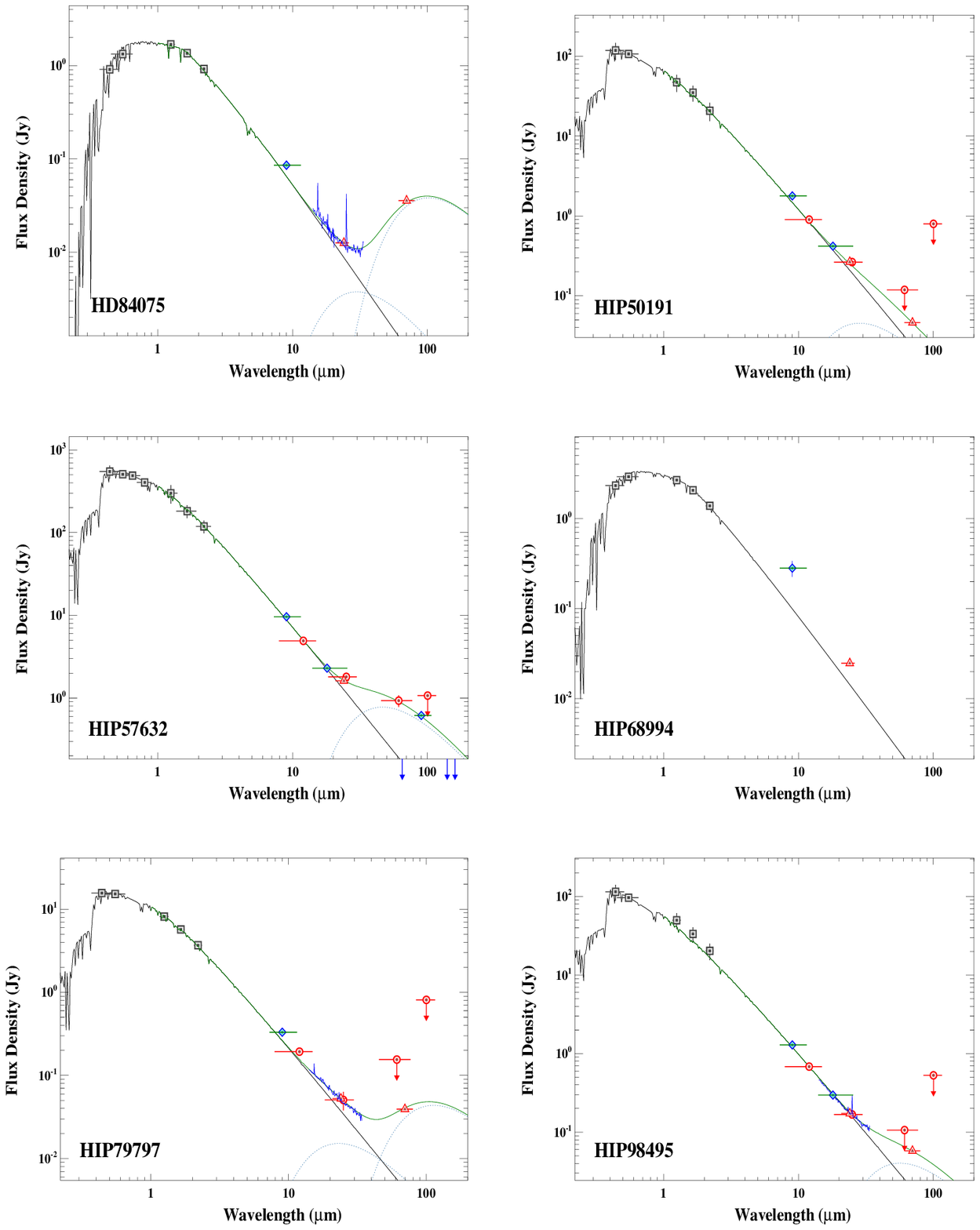}
\caption{\label{figure7} Spectral energy distributions (SEDs) for Argus stars in Table 10 with probable or definite infrared excess emission.}
\end{figure}

\clearpage
\begin{figure}
\includegraphics[width=140mm]{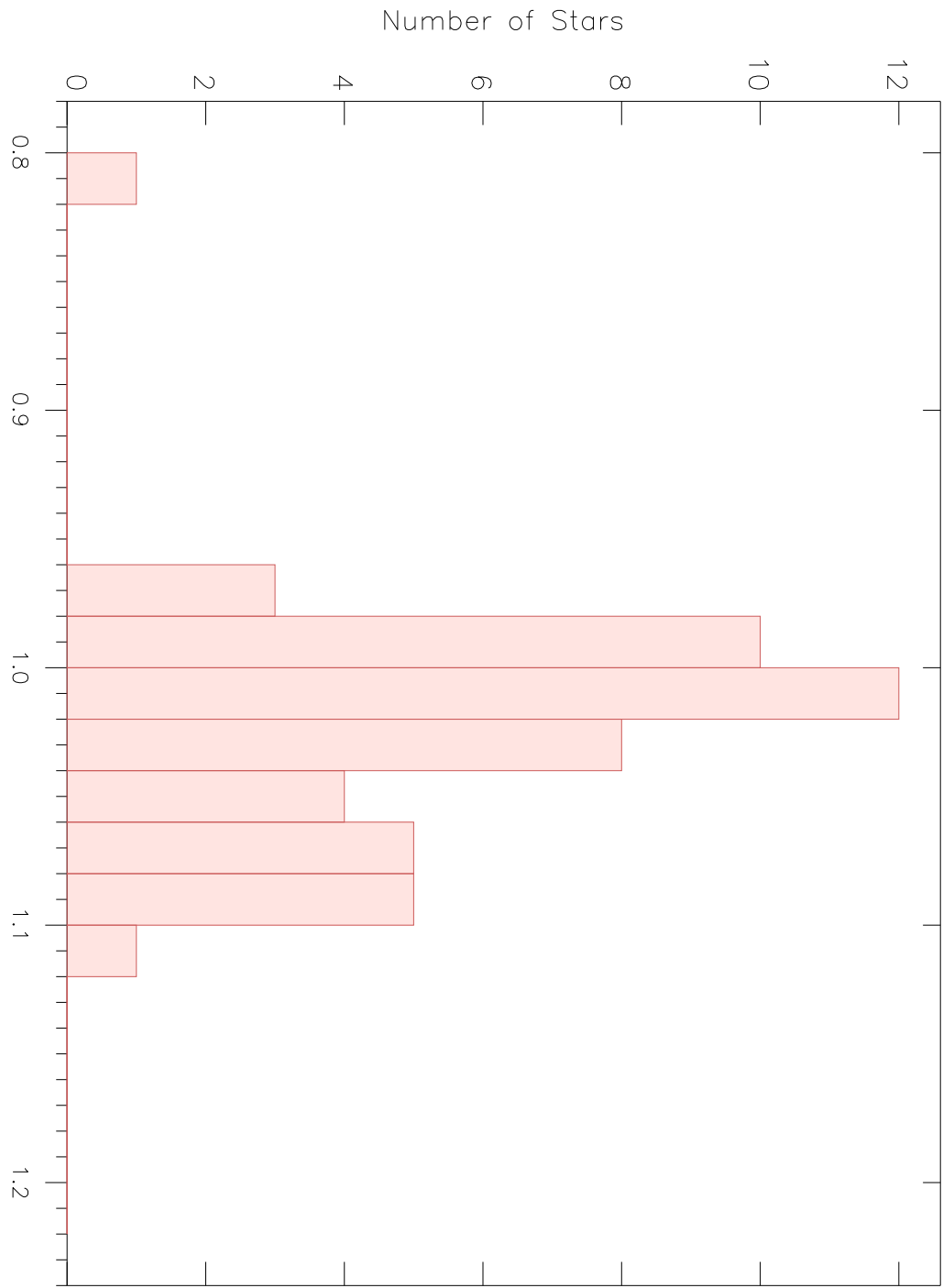}
\caption{\label{figure 8} Histogram of the ratio of MIPS measured to model photospheric 24 $\mu$m flux densities for a sample of the fainter FGK type stars from Trilling et al. (2008) without known MIPS measured excess emission.  See Section 3 for details.}
\end{figure}

\end{document}